\documentclass[showpacs,amsmath,amssymb,prd,nofootinbib,twocolumn]{revtex4}
\pdfoutput=1
\usepackage{graphicx}
\usepackage{dcolumn}
\usepackage{bm}
\usepackage{epsf}
\usepackage{hyperref}
\usepackage{amsfonts}
\usepackage{verbatim}
 
\begin{document}
 
\title{Analytical Tendex and Vortex Fields for Perturbative Black Hole Initial Data}

\author{Kenneth A. Dennison}

\affiliation{Department of Physics and Astronomy, Bowdoin College,
Brunswick, Maine 04011}

\author{Thomas W. Baumgarte}

\affiliation{Department of Physics and Astronomy, Bowdoin College,
Brunswick, Maine 04011}



\begin{abstract}

Tendex and vortex fields, defined by the eigenvectors and eigenvalues of the electric and magnetic parts of the Weyl curvature tensor, form the basis of a recently developed approach to visualizing spacetime curvature.   In particular, this method has been proposed as a tool for interpreting results from numerical binary black hole simulations, providing a deeper insight into the physical processes governing the merger of black holes and the emission of gravitational radiation.  Here we apply this approach to approximate but analytical initial data for both single boosted and binary black holes.  These perturbative data become exact in the limit of small boost or large binary separation.  We hope that these calculations will provide additional insight into the properties of tendex and vortex fields, and will form a useful test for future numerical calculations. 
\end{abstract}


 
\maketitle


\section{Introduction}
\label{INTRO}

The first dynamical simulations of the inspiral and merger of binary black holes \cite{Pre05b,CamLMZ06,BakCCKM06} marked a significant breakthrough in the field of numerical relativity and an important step towards understanding the two-body problem in general relativity.  They also produced the first reliable predictions of the gravitational wave signals emitted in these events, which are needed for the analysis of data from gravitational wave interferometers.  Soon after these initial calculations, which adopted equal-mass and non-spinning binaries, simulations of binaries with unequal masses or non-zero spin revealed important phenomena, including orbital hang-up (see, e.g., \cite{CamLZ06b}), spin-flip (e.g.~\cite{CamLZKM07}) and black hole recoil (e.g.~\cite{HerHHSL06,BakCCKMM06,CamLZM07,GonHSBH07,LouZ11}; see also \cite{BauS10} for a review).

In an effort to gain a better physical understanding of these phenomena, several researchers have developed tools for the visualization and interpretation of gravitational fields and spacetime dynamics.  Here we focus on tendex and vortex fields (see \cite{OweBCKLMNSZZT11,NicOZZBCKLMST11,ZimNZ11}), which are defined in terms of the eigenvectors and eigenvalues of the electric and magnetic parts of the Weyl curvature tensor (see \cite{RezMJ10,JarMMR12a,JarMMR12b} for an alternative ``cross-correlation" approach).  As we will explain in more detail below, the tendex fields describe tidal stretching or compression, while the vortex fields describe precession.  

Tendex and vortex fields and their properties have already been explored for a number of different types of spacetimes.  In \cite{NicOZZBCKLMST11}, the authors focus on applications to weak-field systems and gravitational wave generation by such systems, including the case of a Newtonian slow-motion binary.  Several numerical simulations of binary black hole systems are discussed in \cite{OweBCKLMNSZZT11}, where the new tools are applied to analyze an ``extreme-kick'' merger.  In \cite{ZimNZ11}, the authors use tendex and vortex fields in conjunction with results from topology to investigate asymptotic properties of gravitational radiation.  

In this paper we add to this list an analysis of tendex and vortex fields for approximate but analytical initial data describing boosted and binary black holes (see \cite{DenBP06}).  The initial data are derived as perturbations of Schwarzschild black holes, and become exact in the limit of small boost or large binary separation.  Given that the initial data are analytical, we can also find expressions for the tendex and vortex fields, providing an analytical example of these fields for a strong-field binary.  

Our paper is organized as follows.  In Sec.~\ref{TENDEXVORTEXTOOLS} we briefly review the notion of tendex and vortex fields, following \cite{OweBCKLMNSZZT11}, and apply these tools to a Schwarzschild black hole.  We review the perturbative initial data for single boosted and binary black holes in Sec.~\ref{INITIALDATA}.   In Secs.~\ref{RESULTS_SINGLEBOOSTED} and \ref{RESULTS_BINARY} we apply the new visualization tools to these perturbative initial data, relegating some details of the calculations to several appendices.  We conclude with a brief discussion and summary in Sec.~\ref{SUM}.   Throughout this paper we use geometrical units with $G=c=1$.


\section{Tendex and Vortex Fields}
\label{TENDEXVORTEXTOOLS}

Owen et al.~\cite{OweBCKLMNSZZT11} introduced several tools for visualizing spacetime curvature, based on the eigenvectors and eigenvalues of the electric and magnetic parts of the Weyl curvature tensor.  In Sec.~\ref{DEFINITIONS} we follow \cite{OweBCKLMNSZZT11,NicOZZBCKLMST11,ZimNZ11} and define tendex and vortex fields as well as related quantities needed for the rest of the paper.  In Sec.~\ref{RESULTS_SCHWARZSCHILD} we apply these definitions to a Schwarzschild black hole, which serves both as a pedagogical example and as the background solution in later parts of this paper.

\subsection{Definitions}
\label{DEFINITIONS}

Assuming maximal slicing and vacuum, as we will throughout this paper, we can write the electric and magnetic parts of the Weyl curvature tensor as
\begin{equation}
\label{E_ij_Defined}
{\mathcal E}_{ij}=R_{ij}-K_{i}^{\phantom{i}k}K_{jk}
\end{equation}
and
\begin{equation}
\label{B_ij_Defined}
{\mathcal B}_{ij}=\epsilon_{j}^{\phantom{j}lk}D_{k}K_{li},
\end{equation}
respectively.  Here lowercase Latin indices run over the three spatial coordinates, $R_{ij}$ is the three-dimensional Ricci tensor, $K_{ij}$ is the extrinsic curvature, $D_{k}$ denotes the covariant derivative compatible with the spatial metric $\gamma_{ij}$,  $\epsilon_{ijk}$ is the spatial Levi-Civita tensor, 
\begin{equation}
\epsilon_{ijk} = \gamma^{1/2} [ijk]
\end{equation}
where $[ijk]$ is the alternating symbol with $[123]=+1$, and $\gamma$ is the determinant of the metric.  Maximal slicing implies that the trace of the extrinsic curvature vanishes, $K \equiv \gamma^{ij} K_{ij} = 0$.

Evaluating the electric part of the Weyl tensor on the horizon of a black hole and contracting it twice with the horizon's inward unit normal vector $N^{i}$ yields the {\em horizon tendicity} 
\begin{equation}
\label{ENN_Defined}
{\mathcal E}_{NN} = {\mathcal E}_{ij}N^{i}N^{j}.
\end{equation}
The horizon tendicity measures the strength of the tidal acceleration at the horizon \cite{OweBCKLMNSZZT11}.  The analogous quantity for the magnetic part of the Weyl tensor yields the {\em horizon vorticity} 
\begin{equation}
\label{BNN_Defined}
{\mathcal B}_{NN} = {\mathcal B}_{ij}N^{i}N^{j},
\end{equation}
which measures  the strength of the frame-drag angular acceleration at the horizon \cite{OweBCKLMNSZZT11}.   Horizons can be colored according to the horizon tendicity or vorticity to help illustrate their gravitational properties.  We provide an example in Fig.~\ref{fig1} below, but also refer to the many examples in Refs.~\cite{OweBCKLMNSZZT11,NicOZZBCKLMST11,ZimNZ11}. 

Both ${\mathcal E}_{ij}$ and ${\mathcal B}_{ij}$ are symmetric, and therefore can be characterized by their three orthonormal eigenvectors and the associated eigenvalues.  It is convenient to work in an orthonormal basis where the spatial metric is $\delta_{\hat{\imath}\hat{\jmath}}$.  In such a frame, which we will denote with hats, we do not need to distinguish between contravariant and covariant indices.  The eigenvectors and eigenvalues satisfy familiar equations of the form 
\begin{equation}
\label{Eigenvalue_Equation_Sample}
{\mathcal E}_{\phantom{\hat{\imath}}\hat{\jmath}}^{\hat{\imath}}v^{\hat{\jmath}}=\lambda v^{\hat{\imath}},
\end{equation}
which is equation (3) of \cite{ZimNZ11} except that we raised the free index using the spatial metric.  We then refer to the eigenvectors of ${\mathcal E}_{\hat{\imath}\hat{\jmath}}$ as the {\em tendex fields} and the corresponding eigenvalues as {\em tendicities}, while the corresponding quantities for ${\mathcal B}_{\hat{\imath}\hat{\jmath}}$ are called the {\em vortex fields} and {\em vorticities}.  Finally, we refer to the integral curves of the tendex and vortex fields as {\em tendex lines} and {\em vortex lines}.   Both ${\mathcal E}_{ij}$ and ${\mathcal B}_{ij}$ are also traceless, so that, in an orthonormal basis, their eigenvalues have to add to zero.   In the following we will derive ${\mathcal E}_{ij}$ and ${\mathcal B}_{ij}$ in a coordinate basis and then switch to an orthonormal basis to solve the eigenvalue problem.  

An observer oriented along a tendex vector will tend to be tidally stretched for negative tendicity and tidally compressed for positive tendicity \cite{OweBCKLMNSZZT11,NicOZZBCKLMST11}.  Neighboring gyroscopes oriented along a vortex vector will exhibit counterclockwise differential precession for negative vorticity and clockwise differential precession for positive vorticity \cite{OweBCKLMNSZZT11,NicOZZBCKLMST11}.   In the following Sections we will represent tendex and vortex fields in two-dimensional plots with the help of ``iron filings" that are familiar from representations of magnetic field lines.  The iron filings show the direction of the eigenvector; we simultaneously shade the plot backgrounds according to the corresponding eigenvalues (see \cite{OweBCKLMNSZZT11,NicOZZBCKLMST11,ZimNZ11} for many examples of alternative representations of these fields).

\subsection{A Schwarzschild Black Hole}
\label{RESULTS_SCHWARZSCHILD}

In this Section we compute tendicities and vorticities for a Schwarzschild black hole in isotropic spatial coordinates, both as a pedagogical example and so that we can use the results as the background solution in our later perturbative treatment.  

Consider a Schwarzschild black hole with bare mass ${\mathcal M}$.  For a Schwarzschild black hole the bare mass is equal to the black hole's ADM energy $M_{\rm ADM}$ or the irreducible mass $M_{\rm irr}$.  In this Section we may therefore replace ${\mathcal M}$ with either $M_{\rm ADM}$ or $M_{\rm irr}$.   We write the spatial metric as
\begin{equation}
\label{spatial_metric_conformal_form}
\gamma_{ij}=\psi^{4}\bar{\gamma}_{ij},
\end{equation}
where $\psi$ is the conformal factor and $\bar \gamma_{ij}$ is the conformally related metric.  In isotropic coordinates, on a slice of constant Schwarzschild time $t$, the conformal factor for a Schwarzschild spacetime then takes the form
\begin{equation}
\label{conformal_factor_Schwarzschild_isotropic_spc}
\psi = 1 + \frac{{\mathcal M}}{2r},
\end{equation}
and the conformally related metric is flat; in spherical polar coordinates we have
\begin{equation}
\label{flat_metric_spc}
\bar{\gamma}_{ij}= \mbox{diag}(1,r^2,r^2\sin^2\theta).
\end{equation}
Also, on a slice of constant Schwarzschild time $t$ the extrinsic curvature vanishes identically,
\begin{equation}
\label{extrinsic_curvature_Schwarschild_isotropic}
K_{ij} = 0.
\end{equation}
From (\ref{B_ij_Defined}) we see that the magnetic part of the Weyl tensor is also zero,
\begin{equation}
\label{Bij_Schwarzschild_Is_0}
{\mathcal B}_{ij}=0.
\end{equation}
As one might expect for a non-rotating, static black hole, all vorticities therefore vanish identically.

The tendicities of a Schwarzschild black hole, on the other hand, are non-zero.  In order to evaluate them, we first note that for a vanishing extrinsic curvature, the electric part the Weyl tensor (\ref{E_ij_Defined}) reduces to the Ricci tensor
\begin{equation}
{\mathcal E}_{ij}=R_{ij}.
\end{equation} 
We can evaluate the Ricci tensor using 
\begin{eqnarray}
\label{R_ij_in_conformal_terms}
R_{ij} &=& \bar{R}_{ij}-2(\bar{D}_{i}\bar{D}_{j}\ln\psi+\bar{\gamma}_{ij}\bar{\gamma}^{lm}\bar{D}_{l}\bar{D}_{m}\ln\psi)
\nonumber\\&& +\:4\Big((\bar{D}_{i}\ln\psi)(\bar{D}_{j}\ln\psi)-
\nonumber\\&& \bar{\gamma}_{ij}\bar{\gamma}^{lm}(\bar{D}_{l}\ln\psi)(\bar{D}_{m}\ln\psi)\Big)
\end{eqnarray}
(see, e.g.,  \cite{BauS10}), where $\bar R_{ij}$ and $\bar D_i$ are the Ricci tensor and covariant derivative associated with the conformally related metric $\bar \gamma_{ij}$.  Since the latter is flat, we have $\bar{R}_{ij}=0$, and we find that the only non-zero components of  ${\mathcal E}_{ij}$ are
\begin{equation}
\label{E_rr_appcalc}
{\mathcal E}_{rr} =-\frac{2{\mathcal M}}{\psi^{2}r^{3}}
\end{equation}
and \begin{equation}
\label{E_thetatheta_appcalc}
{\mathcal E}_{\theta\theta}=\frac{{\mathcal E}_{\phi\phi}}{\sin^2\theta}=\frac{{\mathcal M}}{\psi^{2}r}.
\end{equation}

To compute the horizon tendicity, we first note that the inward normal on a sphere of constant radius $r$ is
\begin{equation}
\label{Schwarzschild_inward_normal}
N^{i}=\left(-\frac{1}{\psi^{2}},0,0\right).
\end{equation}
We then have
\begin{equation}
\label{E_NN_appcalc}
{\mathcal E}_{NN} = {\mathcal E}_{ij}N^{i}N^{j} = {\mathcal E}_{rr}N^{r}N^{r} = -\frac{2 {\mathcal M}}{\psi^{6}r^{3}}, 
\end{equation}
which, when evaluated at the horizon $r={\mathcal M}/2$, yields
\begin{equation}
\label{E_NN_appcalc_eval}
{\mathcal E}_{NN} = -\frac{1}{4{\mathcal M}^{2}} = -\frac{^{(2)}\mathcal R}{2},
\end{equation}
where $^{(2)}{\mathcal R}$ is the two-dimensional Ricci scalar for the horizon and the last equality is a consequence of a more general result for quiescent black holes given in Ref. \cite{OweBCKLMNSZZT11}.

To solve the eigenvalue problem in equation (\ref{Eigenvalue_Equation_Sample}), we convert ${\mathcal E}_{ij}$ to a spherical polar orthonormal basis.  The same transformation to an orthonormal basis will be used in later sections - only the conformal factor will be different when we consider perturbed data.  Noting that the spatial metric is defined by $\gamma_{ij}={\mathbf e}_{i}\cdot{\mathbf e}_{j}$,  we now define orthonormal basis vectors 
\begin{equation}
{\mathbf e}_{\hat{r}}=\frac{1}{\psi^2} {\mathbf e}_{r}, ~~~
{\mathbf e}_{\hat{\theta}}= \frac{1}{\psi^{2}r} {\mathbf e}_{\theta}, ~~~
{\mathbf e}_{\hat{\phi}}= \frac{1}{\psi^2r \sin \theta} {\mathbf e}_{\phi},
\end{equation}
with a corresponding dual basis of 1-forms
\begin{equation}
\tilde\omega^{\hat{r}}=\psi^{2}\widetilde{{\mathbf d}r}, ~~~
\tilde\omega^{\hat{\theta}}=\psi^{2}r\widetilde{{\mathbf d}\theta}, ~~~
\tilde\omega^{\hat{\phi}}=\psi^{2}r(\sin\theta)\widetilde{{\mathbf d}\phi}.
\end{equation}
The orthonormal components ${\mathcal E}_{\hat{\imath}\hat{\jmath}}$ can be identified using
\begin{equation}
{\mathcal E}_{ij}\widetilde{dx}^{i}\widetilde{dx}^{j} = {\mathcal E}_{\hat{\imath}\hat{\jmath}}\tilde{\omega}^{\hat{\imath}}\tilde{\omega}^{\hat{\jmath}}.
\end{equation}
For example,
\begin{equation}
\label{electric_convert_rphi}
{\mathcal E}_{\hat{r}\hat{\phi}}= \frac{1}{\psi^4r\sin\theta} {\mathcal E}_{r\phi}.
\end{equation}
The other components of ${\mathcal E}_{\hat{\imath}\hat{\jmath}}$ are similar, and ${\mathcal B}_{ij}$ transforms just like ${\mathcal E}_{ij}$. Applying these transformations to (\ref{E_rr_appcalc}) and (\ref{E_thetatheta_appcalc}) we find that the non-zero components of ${\mathcal E}_{\hat \imath \hat \jmath}$ in a spherical polar orthonormal basis are
\begin{equation}
\label{Err_Schwarzschild_orthonormal}
{\mathcal E}_{\hat{r}\hat{r}} = -2{\mathcal E}_{\hat{\theta}\hat{\theta}} = -2{\mathcal E}_{\hat{\phi}\hat{\phi}} = -\frac{2{\mathcal M}}{\psi^{6}r^{3}}.
\end{equation}
This is consistent with the results in \cite{NicOZZBCKLMST11} and \cite{MisTW73}, but expressed in a different coordinate system.  Equation (\ref{Err_Schwarzschild_orthonormal}) shows that ${\mathcal E}_{\hat \imath \hat \jmath}$ is diagonal in this basis, so its orthonormal eigenvectors can be chosen to be 
\begin{equation}
\label{singleboostedvE1back}
v_{E1}^{(0)\:\hat{\imath}} = (e_{\hat r})^{\hat \imath} = (1,0,0), 
\end{equation}
\begin{equation}
\label{singleboostedvE2back}
v_{E2}^{(0)\:\hat{\imath}} = (e_{\hat \theta})^{\hat \imath} = (0,1,0), 
\end{equation}
and
\begin{equation}
\label{singleboostedvE3back}
v_{E3}^{(0)\:\hat{\imath}} = (e_{\hat \phi})^{\hat \imath} = (0,0,1). 
\end{equation}
Here the superscript ${}^{(0)}$ has been added for consistency with later Sections where these results will serve as background solutions to a perturbative treatment, and the labels $E1$, $E2$ and $E3$ denote these as the three eigenvectors of ${\mathcal E}_{\hat \imath \hat \jmath}$.  The corresponding eigenvalues are given by
\begin{equation}
\label{singleboostedlambdaE1back}
\lambda_{E1}^{(0)} = {\mathcal E}_{\hat{r}\hat{r}}^{(0)} = -\frac{2{\mathcal M}}{\psi_{(0)}^{6}r^{3}} = -\frac{128{\mathcal M}r^{3}}{({\mathcal M}+2r)^{6}}
\end{equation}
and
\begin{equation}
\label{singleboostedlambdaE2E3back}
\lambda_{E2}^{(0)} = \lambda_{E3}^{(0)} = {\mathcal E}_{\hat{\theta}\hat{\theta}}^{(0)}  = \frac{{\mathcal M}}{\psi_{(0)}^{6}r^{3}} = \frac{64{\mathcal M}r^{3}}{({\mathcal M}+2r)^{6}}.
\end{equation}
Since $\lambda_{E1}^{(0)}$ is negative, observers are stretched in the radial direction.   The eigenvalues $\lambda_{E2}^{(0)}$ and $\lambda_{E3}^{(0)}$ are positive and degenerate, so observers are compressed equally in all tangential directions (see also Fig.~2 in either of Refs. \cite{OweBCKLMNSZZT11,NicOZZBCKLMST11} and the associated discussion).
  

\section{Perturbative Black Hole Initial Data}
\label{INITIALDATA}

Numerical relativity simulations using a ``3+1'' decomposition require initial data, namely a spatial metric  $\gamma_{ij}$ and extrinsic curvature $K_{ij}$ satisfying the Hamiltonian constraint and the momentum constraint.  Using the conformal transformation (\ref{spatial_metric_conformal_form}), and transforming the extrinsic curvature according to
\begin{equation}
\label{Kij_in_terms_of_Aij}
K_{ij} = \psi^{-2}\bar{A}_{ij},
\end{equation}
the Hamiltonian constraint takes the form
\begin{equation}
\label{Hamiltonian_Constraint}
\bar{D}^{2}\psi = -\frac{1}{8}\psi^{-7}\bar{A}_{ij}\bar{A}^{ij},
\end{equation}
while the momentum constraint reduces to
\begin{equation}
\label{Momentum_Constraint}
\bar{D}_{j}\bar{A}^{ij} = 0.
\end{equation}
Here $\bar{D}_{i}$ denotes the covariant derivative associated with the conformally related metric $\bar \gamma_{ij}$, and $\bar D^2 \equiv \bar \gamma^{ij} \bar D_i \bar D_j$ is the associated Laplace operator.  In the above expressions we have again assumed maximal slicing and vacuum; we have also assumed conformal flatness, meaning that $\bar \gamma_{ij}$ is a flat metric and $\bar D^2$ a flat-space Laplace operator. 

The momentum constraint has become linear under these assumptions, and is solved analytically by so-called ``Bowen-York" solutions \cite{Bow79,BowY80,Yor89}.  For a black hole at coordinate location $C^{i}$ with linear momentum $P^{i}$ these solutions are given by 
\begin{equation}
\label{Aijup}
\bar{A}^{ij}_{\rm{CP}} = \frac{3}{2r^{2}_{\rm{C}}}\left[P^{i}n^{j}_{\rm{C}}+P^{j}n^{i}_{\rm{C}}-
(\bar{\gamma}^{ij}-n^{i}_{\rm{C}}n^{j}_{\rm{C}})P_{k}n^{k}_{\rm{C}}\right],
\end{equation}
where $r_{\rm{C}} = |x^{i}-C^{i}|$ is the coordinate distance from the center of
the black hole and $n^{i}_{\rm{C}} = (x^{i}-C^{i})/r_{\rm{C}}$ is the unit vector (normalized with respect to the conformally-related background metric) which points from the center to coordinate location $x^{i}$. 

The solutions (\ref{Aijup}) can then be inserted into the Hamiltonian constraint (\ref{Hamiltonian_Constraint}), which, in general, still has to be solved numerically (see, e.g., \cite{KulSY83,CooY90,Coo94,BraB97,Bau00,Bau12} for different approaches and results, as well as \cite{BauS10} for a review.)  Here we will review a perturbative but analytical approach (see \cite{GleKP02,Lag04,DenBP06}).  We note that for  vanishing boost, the Hamiltonian constraint (\ref{Hamiltonian_Constraint}) is solved exactly by the Schwarzschild conformal factor (\ref{conformal_factor_Schwarzschild_isotropic_spc}).  For non-zero boost, we can then consider the leading-order perturbations of the Schwarzschild conformal factor.  In the following two Sections we will consider single boosted black holes and binary black holes separately.


\subsection{Single Boosted Black Holes}
\label{ID_SINGLEBOOSTED}

We first note that $\bar A^{ij}$ is linear in the magnitude $P$ of the momentum.  From the Hamiltonian constraint (\ref{Hamiltonian_Constraint}) we then see that all perturbations of $\psi$ can only contain even powers of $P$.   Defining 
\begin{equation}
\epsilon_P \equiv \frac{P}{\mathcal M},
\end{equation}
we can therefore write the solution to the Hamiltonian constraint (\ref{Hamiltonian_Constraint}) as 
\begin{equation}
\label{conformal_single_boosted_short}
\psi = \psi_{(0)} + \epsilon_{P}^{2}u + {\mathcal O}(\epsilon_{P}^{4}),
\end{equation}
where $\psi_{(0)}$ is the Schwarzschild conformal factor (\ref{conformal_factor_Schwarzschild_isotropic_spc}) for a black hole at coordinate location $C^i$,
\begin{equation}
\label{define_psi0}
\psi_{(0)} = 1 + \frac{{\mathcal M}}{2r_{C}},
\end{equation}
and where the function $u$ can be written as 
\begin{equation}
\label{pertboostusol}
u = \frac{{\mathcal M}}
{8({\mathcal M}+2r_{\rm{C}})^{5}}
\Big( u_{0}(r_{\rm{C}}) P_{0}(\cos\theta)
+ u_{2}(r_{\rm{C}}) P_{2}(\cos\theta) \Big) .
\end{equation}
Here
\begin{equation}
\label{Legendre_P_0_equation}
P_{0}(\cos\theta)=1
\end{equation}
and
\begin{equation}
\label{Legendre_P_2_equation}
P_{2}(\cos\theta)=\frac{3}{2}(\cos^{2}\theta)-\frac{1}{2}
\end{equation}
are Legendre polynomials with $\theta$ measured from the boost direction, and
the radial functions 
$u_{0}(r_{\rm{C}})$ and $u_{2}(r_{\rm{C}})$
appearing in equation (\ref{pertboostusol}) are given by
\begin{equation}
\label{pertboostu_0}
u_{0}(r_{\rm{C}})={\mathcal M}^{4}+
10{\mathcal M}^{3}r_{\rm{C}}+40{\mathcal M}^{2}r_{\rm{C}}^{2}+
80{\mathcal M}r_{\rm{C}}^{3}+80r_{\rm{C}}^{4},
\end{equation}
and
\begin{eqnarray}
\label{pertboostu_2}
u_{2}(r_{\rm{C}})&=&
\frac{{\mathcal M}}{5r^{3}_{\rm{C}}}
\Bigg( 42{\mathcal M}^{5}r_{\rm{C}}+378{\mathcal M}^{4}r_{\rm{C}}^{2}+
1316{\mathcal M}^{3}r_{\rm{C}}^{3} \nonumber \\ 
&& +\:2156{\mathcal M}^{2}r_{\rm{C}}^{4}+
1536{\mathcal M}r_{\rm{C}}^{5}+ 240r_{\rm{C}}^{6}\nonumber \\
& &
+\:21{\mathcal M}\left({\mathcal M}+2r_{\rm{C}}\right)^{5}
\ln\left(\frac{{\mathcal M}}{{\mathcal M}+2r_{\rm{C}}}\right) \Bigg)
\end{eqnarray}
(see \cite{GleKP02,Lag04}).  


\subsection{Binary Black Hole Systems}
\label{ID_BINARY}

The initial data discussed in Sec. \ref{ID_SINGLEBOOSTED} can be used to construct
similar initial data for a binary black hole system with one hole at coordinate location $C_{1}^{i}$ with bare mass ${\mathcal M}_{1}$ and momentum $P_{1}^{i}$, and another at coordinate location $C_{2}^{i}$ with bare mass ${\mathcal M}_{2}$ and momentum $P_{2}^{i}$ (see \cite{DenBP06}). 

The momentum constraint (\ref{Momentum_Constraint}) is linear in $\bar{A}^{ij}$, so it can be solved by adding two solutions of the form shown in equation (\ref{Aijup}):
\begin{equation}
\label{Aijup_binary}
\bar{A}^{ij} = \bar{A}^{ij}_{\rm{C_{1}P_{1}}} + \bar{A}^{ij}_{\rm{C_{2}P_{2}}}.
\end{equation}
The Hamiltonian constraint (\ref{Hamiltonian_Constraint}) is not linear in $\psi$, but it is still possible to construct a perturbative solution similar to that for a single boosted black hole by adding the separate perturbations from each of the holes.  The derivation in \cite{DenBP06} assumes that the coordinate distance $s$ between the holes obeys
\begin{eqnarray}
\label{lengthscales}
s \gg &{\mathcal M}_{1}& \gg P_{1},\nonumber\\
s \gg &{\mathcal M}_{2}& \gg P_{2},
\end{eqnarray}
with $P_{1}/{\mathcal M}_{1}\sim P_{2}/{\mathcal M}_{2}\sim\epsilon_{P}$.  The conformal factor satisfying equation (\ref{Hamiltonian_Constraint}) to second order in the boosts is
\begin{equation}
\label{pertbinpsisol}
\psi=1+\frac{{\mathcal M}_{1}}{2r_{\rm{C_{1}}}}+
\frac{{\mathcal M}_{2}}{2r_{\rm{C_{2}}}}+\epsilon_{P}^{2}u_{1}+\epsilon_{P}^{2}u_{2}+
{\mathcal O}(\epsilon_{P}^{4}),
\end{equation}
where $u_{1}$ and $u_{2}$ are appropriate versions of equation (\ref{pertboostusol}), with $r_{\rm C}$ replaced by either $r_{\rm C_1}$ or $r_{\rm C_2}$.

In this paper we will assume an equal-mass binary with ${\mathcal M}_{1}={\mathcal M}_{2}={\mathcal M}$.  We also assume a quasicircular orbit, which implies that the black holes have  equal magnitude momenta $P_{1}=P_{2}=P$, so $\epsilon_{P_{1}}=\epsilon_{P_{2}}=\epsilon_{P}$.  Applying the virial relationship from equation (57) of \cite{DenBP06} results in the Kepler law
\begin{equation}
\label{virial}
\frac{{\mathcal M}}{2s} = \epsilon_{P}^{2} + {\mathcal O}(\epsilon_{P}^{4}).
\end{equation}
Note that in using equation (57) of \cite{DenBP06} we have replaced the irreducible mass \cite{Chr70} with the bare mass, but this difference only changes the calculation at order  ${\mathcal O}(\epsilon_{P}^{4})$, so it can be ignored here.  

For both single boosted and binary black holes, $\bar A^{ij}$ is linear in $P$, so that perturbations of $\psi$ contain only even powers of $P$.  From (\ref{Kij_in_terms_of_Aij}) we then see that $K_{ij}$ can only have odd powers of $P$.   Therefore, perturbations of the Schwarzschild expressions for ${\mathcal E}_{ij}$ also contain only even powers of $P$, while perturbations of  ${\mathcal B}_{ij}$ contain only odd powers.


\section{Tendex and vortex fields for boosted black holes}
\label{RESULTS_SINGLEBOOSTED}

We now consider a single black hole at the origin boosted in the $+z$ direction, described by the perturbative initial data from Sec. \ref{ID_SINGLEBOOSTED}.   We then compute the tendex and vortex fields to leading-order, adopting the results of Sec. \ref{RESULTS_SCHWARZSCHILD} as the unperturbed background solution.

\subsection{The Electric Part of the Weyl Tensor}
\label{RESULTS_SINGLEBOOSTED_Eij}

As we have discussed above, only even powers of $\epsilon_{P}$ can appear in perturbations of the electric part of the Weyl tensor, so that we may write the result as 
\begin{equation}
\label{E_ij_pert_almost_general_results}
{\mathcal E}_{\hat{\imath}\hat{\jmath}} = {\mathcal E}_{\hat{\imath}\hat{\jmath}}^{(0)} + \epsilon_{P}^{2}{\mathcal E}_{\hat{\imath}\hat{\jmath}}^{(2)}  +
{\mathcal O}(\epsilon_{P}^{4}).
\end{equation} 
For a black hole at the origin we can replace $r_{C}$ with $r$ and find that in a spherical polar orthonormal basis, ${\mathcal E}_{\hat \imath \hat \jmath}^{(0)}$ has nonzero components given by equation (\ref{Err_Schwarzschild_orthonormal}) with the conformal factor $\psi$ replaced by $\psi_{(0)}$.   The perturbations ${\mathcal E}_{ij}^{(2)}$, expressed in a coordinate basis, are computed in Appendix \ref{APP_CALCS_SINGLEBOOSTED_ELECTRIC} (see equations (\ref{E_rr_P2}-\ref{E_phiphi_P2})).  We then obtain the perturbations of the orthonormal components from
\begin{equation}
\label{E_rr_orthonormal_P2}
{\mathcal E}_{\hat{r}\hat{r}}^{(2)} = \frac{1}{\psi^4_{(0)}} {\mathcal E}_{rr}^{(2)}
- \frac{4u}{\psi^5_{(0)}} {\mathcal E}_{rr}^{(0)},
\end{equation}
\begin{equation}
\label{E_rtheta_orthonormal_P2}
{\mathcal E}_{\hat{r}\hat{\theta}}^{(2)} =   {\mathcal E}_{\hat{\theta}\hat{r}}^{(2)} =  \frac{1}{\psi^4_{(0)} r}{\mathcal E}_{r\theta}^{(2)},
\end{equation}
\begin{equation}
\label{E_thetatheta_orthonormal_P2}
{\mathcal E}_{\hat{\theta}\hat{\theta}}^{(2)} = \frac{1}{\psi^4_{(0)} r^2}  {\mathcal E}_{\theta\theta}^{(2)}- \frac{4u}{\psi^5_{(0)} r^2} {\mathcal E}_{\theta\theta}^{(0)},
\end{equation}
and
\begin{equation}
\label{E_phiphi_orthonormal_P2}
{\mathcal E}_{\hat{\phi}\hat{\phi}}^{(2)} =  \frac{1}{\psi^4_{(0)}r^2 \sin^2 \theta } {\mathcal E}_{\phi\phi}^{(2)} - \frac{4u}{\psi^5_{(0)} r^2 \sin^2 \theta} {\mathcal E}_{\phi\phi}^{(0)},
\end{equation}
where the function $u$ is given by equation (\ref{pertboostusol}).

\subsection{The Magnetic Part of the Weyl Tensor}

Only odd powers of $\epsilon_{P}$ appear in the magnetic part of the Weyl tensor, so, to leading order, it can be written in the form 
\begin{equation}
\label{B_ij_pert_almost_general_results}
{\mathcal B}_{\hat{\imath}\hat{\jmath}} = \epsilon_{P}{\mathcal B}_{\hat{\imath}\hat{\jmath}}^{(1)} + {\mathcal O}(\epsilon_{P}^{3}).
\end{equation}
In Appendix \ref{APP_CALCS_SINGLEBOOSTED_MAGNETIC} we show that the only non-vanishing components of the leading-order term ${\mathcal B}_{ij}^{(1)}$, expressed in a coordinate basis, are the $r\phi$ and $\phi r$ components
\begin{equation} \label{B_rphi_singleboosted}
{\mathcal B}_{r \phi}^{(1)} = {\mathcal B}_{\phi r}^{(1)}
= - \frac{96 {\mathcal M} r^3 \sin^2 \theta}{({\mathcal M} + 2 r)^5}.
\end{equation}
We then use the transformation (\ref{electric_convert_rphi}) to find the orthonormal components 
\begin{equation}
\label{B_rphi_orthonormal_P}
{\mathcal B}_{\hat{r}\hat{\phi}}^{(1)} = {\mathcal B}_{\hat{\phi}\hat{r}}^{(1)} = 
\frac{1}{ \psi^4_{(0)} r \sin \theta} {\mathcal B}_{r\phi}^{(1)}.
\end{equation}

\subsection{Horizon Tendicity and Vorticity}
\label{RESULTS_SINGLEBOOSTED_HORIZON}

In Appendix \ref{APP_CALCS_SINGLEBOOSTED_HORIZON} we adopt the results for ${\mathcal E}_{ij}$ in a coordinate basis to find the horizon tendicity.  This calculation also relies on results for the perturbed horizon location that were found in \cite{DenBP06}.  We compute the horizon tendicity directly, following the procedure outlined in Sec. \ref{DEFINITIONS}, but in an alternative approach it is also possible to employ the Newman-Penrose formalism \cite{NewP62} (see \cite{NicOZZBCKLMST11}).  The result of this calculation is
\begin{eqnarray}
{\mathcal E}_{NN} &=& -\frac{1}{4{\mathcal M}^{2}}+\epsilon_{P}^{2} \frac{1}{16{\mathcal M}^{2}} + \\ 
& & \epsilon_{P}^{2} \frac{(-1871+2688\ln 2)}{640 {\mathcal M}^{2}}P_{2}(\cos\theta) 
+ {\mathcal O}(\epsilon_{P}^{4}). \nonumber 
\end{eqnarray}
The first two terms simplify if we express results in terms of the black hole's irreducible mass.  From equation (26) of \cite{DenBP06} we have\footnote{We note that this result holds only for wormhole data.  For trumpet data, for example, the bare mass ${\mathcal M}$ appears to be equal to the irreducible mass $M_{\rm irr}$ for single boosted black holes; see \cite{ImmB09,HanHO09}.}
\begin{equation}
{\mathcal M} = M_{\rm irr} \left( 1-\frac{P^{2}}{8M_{\rm irr}^{2}} \right) + {\mathcal O}(\epsilon_{P}^{4}),
\end{equation}
Inserting this relation into the expression above we see that the (proper area weighted) average value of the horizon tendicity is
\begin{equation}
{\mathcal E}_{NN}^{\rm ave} = - \frac{1}{4 M_{\rm irr}^2} + {\mathcal O}(\epsilon_{P}^{4}).
\end{equation}
The simplicity of this result is not surprising on physical grounds and we might have anticipated it from the Gauss-Bonnet theorem as discussed in this context in \cite{OweBCKLMNSZZT11}.  As shown in Appendix \ref{APP_CALCS_SINGLEBOOSTED_HORIZON}, the average horizon tendicity for a horizon with spherical topology will take the above form whenever a particular combination of spin coefficients vanishes when integrated over the horizon -- which is true here to the relevant order. 

\begin{figure}
\includegraphics[width=0.48\textwidth]{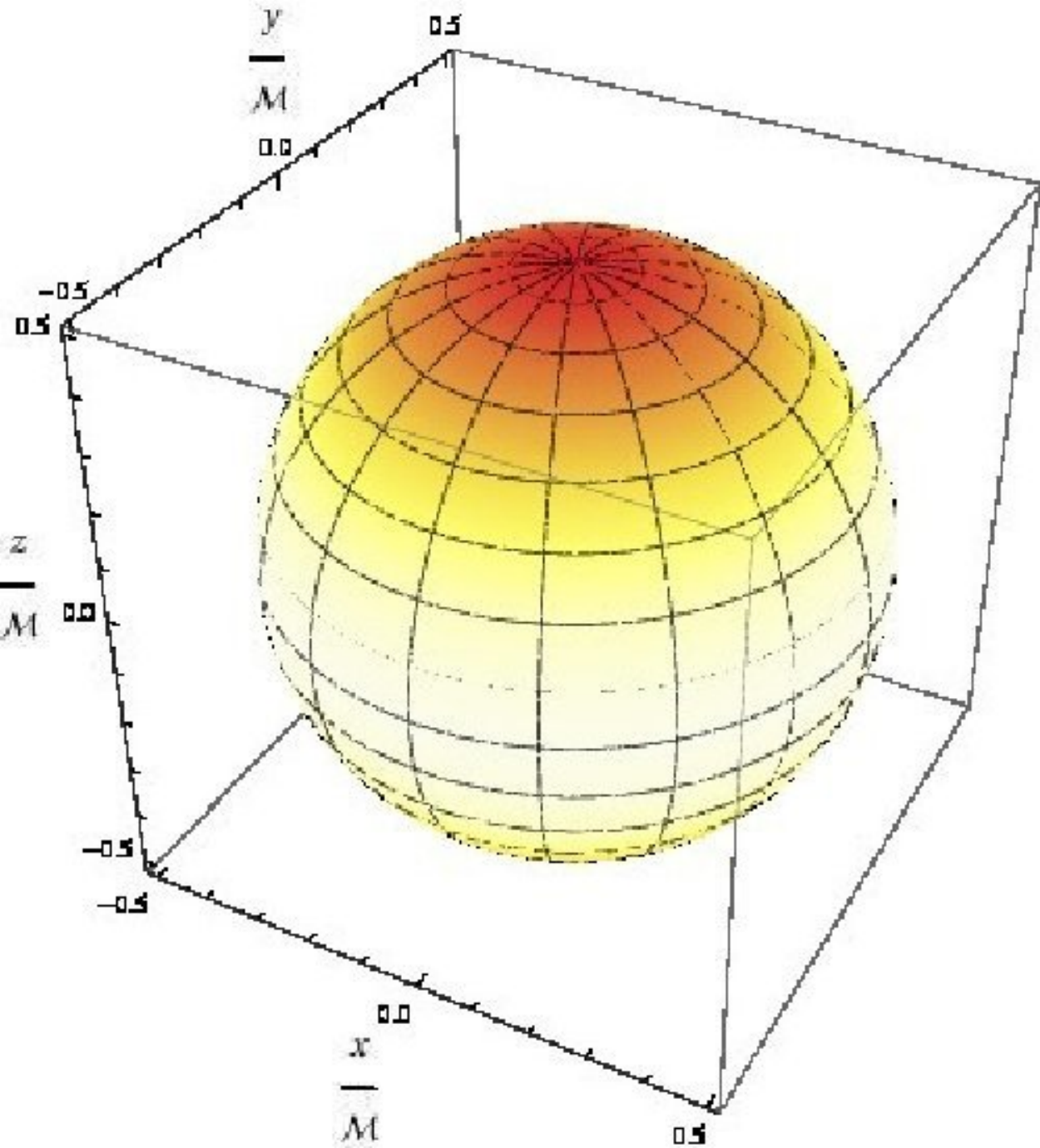}
\caption{The deviation from the average horizon tendicity for a single black hole boosted with $P = 0.1 {\mathcal M}$ in the positive $z$ direction (i.e.~pointing up).   The horizon tendicity is negative everywhere, but it is more negative (darker shading) near the poles at $\theta=0$ and $\theta=\pi$ and less negative (lighter shading) near the equator at $\theta=\pi/2$.  The true distorted shape of the horizon is shown, but for the small boosts relevant for this paper the distortion from the unperturbed spherical shape is not readily apparent.}
\label{fig1}
\end{figure}

To leading order in $\epsilon_P$, the deviation from this average value is proportional to $P_{2}(\cos\theta)$.  In Fig.~\ref{fig1} we plot this deviation of the horizon tendicity from its average value for a black hole boosted in the $+z$ direction with $P = 0.1 {\mathcal M}$.  

In contrast to the horizon tendicity, the horizon vorticity vanishes, at least to the order of our analysis.  Following the same steps that were used to derive ${\mathcal E}_{NN}$ in equations (\ref{boosted_horizon_normal}) - (\ref{boosted_horizon_tendicity_2}), we see that the horizon vorticity is
\begin{equation}
\label{single_boosted_horizon_vorticity}
{\mathcal B}_{NN} = {\mathcal B}_{rr}^{(1)}\gamma^{rr}_{(0)}+{\mathcal O}(\epsilon_{P}^{3}) = {\mathcal O}(\epsilon_{P}^{3}),
\end{equation}
because from equation (\ref{B_rphi_singleboosted_app}) the only nonzero components of ${\mathcal B}_{ij}$ to this order are ${\mathcal B}_{r\phi}$ and ${\mathcal B}_{\phi r}$.   

\subsection{Tendex Fields}

In order to find the tendex fields we have to find the eigenvectors and eigenvalues of the electric part of the Weyl tensor.  We have already solved this problem for the unboosted background solution in Sec. \ref{RESULTS_SCHWARZSCHILD}.  We can now find the leading-order corrections to the Schwarzschild results by solving the eigenvalue problem perturbatively.   While this ``stationary perturbation theory" technique is well known, especially in the context of quantum mechanics, we summarize the most important results, and specialize to the three-dimensional matrices encountered in this context, in Appendix \ref{APP_QMGR}.  Given that two of the unperturbed eigenvalues are degenerate (see equation (\ref{singleboostedlambdaE2E3back})), both degenerate and nondegenerate perturbation theory are required. 

Applying these techniques to the perturbations of the electric part of the Weyl tensor as derived in Sec. \ref{RESULTS_SINGLEBOOSTED_Eij} above, we find that the nondegenerate eigenvalue (\ref{singleboostedlambdaE1back}) generalizes to
\begin{equation}
\label{lambdaE1pieces}
\lambda_{E1} = \lambda_{E1}^{(0)} + \epsilon_{P}^{2}\lambda_{E1}^{(2)} + {\mathcal O}(\epsilon_{P}^{4}).
\end{equation}
Here the background term $\lambda_{E1}^{(0)}$ is given by equation (\ref{singleboostedlambdaE1back}) and 
\begin{eqnarray}
\lambda_{E1}^{(2)} &=& \frac{4 {\mathcal M}}{5 ({\mathcal M}+2
   r)^{12}}\bigg(-21 {\mathcal M}^2\Big(3 \big(\cos 2 \theta \big)+1\Big) \nonumber\\&&\times\:\Big({\mathcal M}^2+2 {\mathcal M}
   r+24 r^2\Big)\Big({\mathcal M}+2 r\Big)^5 \nonumber\\&&\times\:\ln \Big(\frac{{\mathcal M}}{{\mathcal M}+2 r}\Big)-6 {\mathcal M} r
   \Big(\cos 2 \theta \Big)\times\nonumber\\&&\Big(21 {\mathcal M}^7+231 {\mathcal M}^6 r+1540 {\mathcal M}^5 r^2+6930 {\mathcal M}^4
   r^3\nonumber\\&&+\:18720 {\mathcal M}^3 r^4+27568 {\mathcal M}^2 r^5+18816 {\mathcal M} r^6\nonumber\\&&+\:3360 r^7\Big)-2 r
   \Big(21 {\mathcal M}^8+231 {\mathcal M}^7 r+1550 {\mathcal M}^6 r^2\nonumber\\&&+\:6930 {\mathcal M}^5 r^3+18120 {\mathcal M}^4 r^4+24368
   {\mathcal M}^3 r^5\nonumber\\&&+\:11616 {\mathcal M}^2 r^6-4320 {\mathcal M} r^7+3200 r^8\Big)\bigg).
\end{eqnarray}
The corresponding eigenvector is
\begin{equation}
\label{singleboostedvE1pieces}
v_{E1}^{\hat{\imath}} = v_{E1}^{(0)\:\hat{\imath}} + \epsilon_{P}^{2}v_{E1}^{(2)\:\hat{\imath}} + {\mathcal O}(\epsilon_{P}^{4}),
\end{equation}
where the background term $v_{E1}^{(0)\:\hat{\imath}}$ is given by equation (\ref{singleboostedvE1back}) and where the only non-vanishing component of $v_{E1}^{(2) \: \hat \imath}$ is
\begin{eqnarray}
\label{singleboostedvE1theta}
v_{E1}^{(2)\:\hat{\theta}} &=& -\frac{{\mathcal M}}{80 r^3 ({\mathcal M}+2 r)^6}\Bigg(3840 r^8 \sin \theta  \cos \theta \nonumber\\&&-\bigg(\sin 2 \theta \bigg) \bigg({\mathcal M}+2 r\bigg) \bigg(2 r \Big(21 {\mathcal M}^6+357 {\mathcal M}^5 r\nonumber\\&&+\:2170 {\mathcal M}^4 r^2+6342 {\mathcal M}^3 r^3+9388
   {\mathcal M}^2 r^4\nonumber\\&&+\:6216 {\mathcal M} r^5+720 r^6\Big)+21 {\mathcal M} \Big({\mathcal M}+8 r\Big)\nonumber\\&&\times\:\Big({\mathcal M}+2 r\Big)^5 \ln \Big(\frac{{\mathcal M}}{{\mathcal M}+2 r}\Big)\bigg)\Bigg).
\end{eqnarray}

The degenerate unperturbed eigenvalues become
\begin{equation}
\label{lambdaE2pieces}
\lambda_{E2} = \lambda_{E2}^{(0)} + \epsilon_{P}^{2}\lambda_{E2}^{(2)} + {\mathcal O}(\epsilon_{P}^{4})
\end{equation}
and
\begin{equation}
\label{lambdaE3pieces}
\lambda_{E3} = \lambda_{E3}^{(0)} + \epsilon_{P}^{2}\lambda_{E3}^{(2)} + {\mathcal O}(\epsilon_{P}^{4})
\end{equation}
where the identical background terms  $\lambda_{E2}^{(0)}$ and $\lambda_{E3}^{(0)}$ are given by equation (\ref{singleboostedlambdaE2E3back}).  The two perturbative corrections are
\begin{eqnarray}
\label{lambdaE2pertpieceP}
\lambda_{E2}^{(2)} &=& \frac{4 {\mathcal M}}{5 ({\mathcal M}+2 r)^{12}} \left(3 {\mathcal M}A -2 rB \right)
\end{eqnarray}
and
\begin{eqnarray}
\label{lambdaE3pertpieceP}
\lambda_{E3}^{(2)} &=& \frac{4 {\mathcal M}}{5 ({\mathcal M}+2 r)^{12}} \left(3 {\mathcal M}C +2 rD \right),
\end{eqnarray}
where the coefficients $A$, $B$, $C$ and $D$ are given by
\begin{eqnarray}
A &=& 7 {\mathcal M} \Big({\mathcal M}+2 r\Big)^5\ln
   \Big(\frac{{\mathcal M}}{{\mathcal M}+2 r}\Big) \nonumber\\&&\times\Big(3 \big(\cos 2 \theta \big)\big({\mathcal M}^2+3 {\mathcal M}
   r+14 r^2\big)\nonumber\\&& ~~~~~~~~~~~~~~~~~~~~~~~~~ -\:\big({\mathcal M}-r\big) \big({\mathcal M}+6 r\big)\Big)\nonumber\\&&+\:2 r \Big(\cos 2 \theta \Big) \Big(21
   {\mathcal M}^7+252 {\mathcal M}^6 r+1519 {\mathcal M}^5 r^2\nonumber\\&&+\:5698 {\mathcal M}^4 r^3+13216 {\mathcal M}^3 r^4+17536 {\mathcal M}^2
   r^5\nonumber\\&&+\:11184 {\mathcal M} r^6+2400 r^7\Big),
\end{eqnarray}
\begin{eqnarray}
B &=& 21 {\mathcal M}^8 + 294 {\mathcal M}^7
   r+1472 {\mathcal M}^6 r^2+3234 {\mathcal M}^5 r^3\nonumber\\&&+\:2508 {\mathcal M}^4 r^4-928 {\mathcal M}^3 r^5-480 {\mathcal M}^2 r^6\nonumber\\&&+\:4320
   {\mathcal M} r^7-1600 r^8,
\end{eqnarray}
\begin{eqnarray}
C &=& 7 {\mathcal M} \Big({\mathcal M}+2 r\Big)^5 \ln
   \Big(\frac{{\mathcal M}}{{\mathcal M}+2 r}\Big) \nonumber\\&&\times\:\Big(2 {\mathcal M}^2-3 r \big(\cos 2 \theta \big)\big({\mathcal M}-10r\big)+7 {\mathcal M} r+18 r^2\Big)\nonumber\\&&+\:2 r^2 \Big(\cos 2 \theta \Big)\Big(-21 {\mathcal M}^6+21 {\mathcal M}^5
   r+1232 {\mathcal M}^4 r^2 \nonumber\\&&+\: 5504 {\mathcal M}^3 r^3 + 10032 {\mathcal M}^2 r^4+7632 {\mathcal M} r^5\nonumber\\&&+\:960r^6\Big),
\end{eqnarray}
and
\begin{eqnarray}
D &=& 42 {\mathcal M}^8+525 {\mathcal M}^7 r+3022 {\mathcal M}^6 r^2+10164 {\mathcal M}^5
   r^3\nonumber\\&&+\:20628 {\mathcal M}^4 r^4+23440 {\mathcal M}^3 r^5+11136 {\mathcal M}^2 r^6\nonumber\\&&+\:1600r^8.
\end{eqnarray}
The corresponding eigenvectors are
\begin{equation}
\label{singleboostedvE2}
v_{E2}^{\hat{\imath}} = v_{E2}^{(0) \: \hat \imath} + \epsilon_P^2  v_{E2}^{(2) \: \hat \imath} + {\mathcal O}(\epsilon_{P}^{4})
\end{equation}
and
\begin{equation}
\label{singleboostedvE3}
v_{E3}^{\hat{\imath}} = v_{E3}^{(0) \: \hat \imath} + {\mathcal O}(\epsilon_{P}^{4}),
\end{equation}
where the only nonvanishing component of $v_{E2}^{(2) \: \hat \imath}$ is
\begin{equation}
v_{E2}^{(2) \: \hat r} = - v_{E1}^{(2) \: \hat \theta}. 
\end{equation}
The eigenvectors $v_{E1}^{\hat \imath}$, $v_{E2}^{\hat \imath}$ and $v_{E3}^{\hat \imath}$ 
are orthonormal to second order in $\epsilon_{P}$ as expected.  For small boosts, these eigenvectors are not visibly different from the purely radial or purely tangential unperturbed eigenvectors.  Algebraically, we see that the perturbation mixes the ${\bf e}_{\hat r}$ and ${\bf e}_{\hat \theta}$ eigenvectors of the unperturbed state but leaves the ${\bf e}_{\hat \phi}$ eigenvector unchanged.  This is expected because the boost breaks spherical symmetry but preserves axisymmetry.  

\subsection{Vortex Fields}

\begin{figure}
\includegraphics[width=0.48\textwidth]{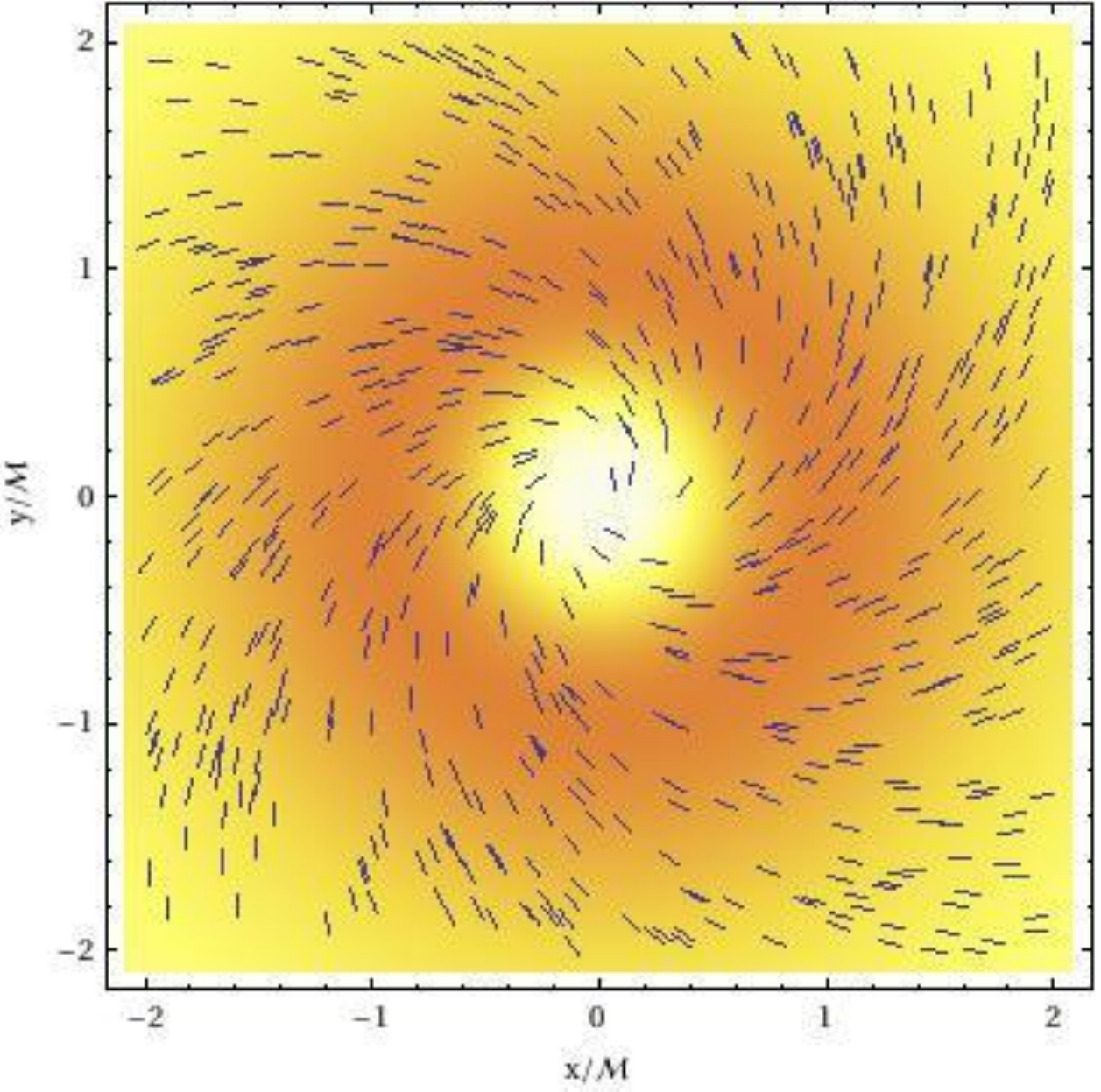}
\caption{The vortex eigenvalue $\lambda_{B2}$ in the $z=0$ plane for a black hole boosted in the positive $z$ direction, along with the projection of eigenvector $v_{B2}^{\hat \imath}$ into that plane.  Randomly placed line segments of fixed length mimic the appearance of ``iron filings" and trace the orientation of the eigenvector, while the colored background shading shows the behavior of the eigenvalue.  The eigenvalue is most negative where the shading is darkest and approaches zero as the shading fades away.  In this figure the black hole is boosted in the direction pointed out of the paper, at the reader.  Note that this figure and later figures bear some resemblance to the ridge patterns in Fig.~3 of \cite{ZimNZ11}.}
\label{figB2z0}
\end{figure}

\begin{figure}
\includegraphics[width=0.48\textwidth]{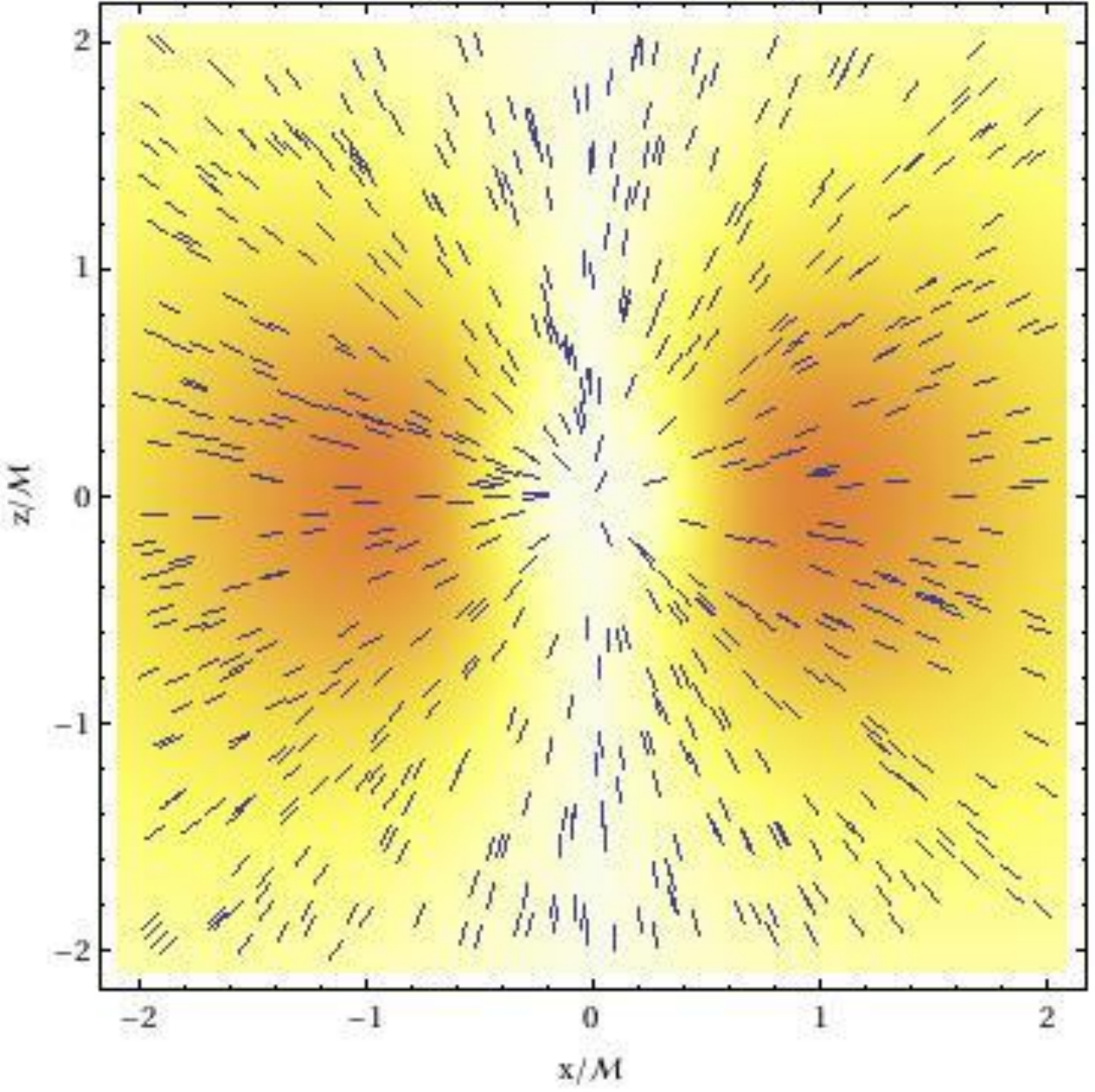}
\caption{The vortex eigenvalue $\lambda_{B2}$ in the $y=0$ plane along with the projection of eigenvector $v_{B2}^{\hat \imath}$ into that plane.  In this figure the boost of the black hole is pointing upwards.}
\label{figB2y0}
\end{figure}

As we have seen above, only odd powers of $P$ enter into an expansion of the magnetic part of the Weyl tensor.  For Schwarzschild we have ${\mathcal B}_{\hat{\imath}\hat{\jmath}}=0$, so even for small boosts we cannot treat this problem perturbatively.  To leading order, ${\mathcal B}_{\hat{\imath}\hat{\jmath}}$ takes the simple form given by equations (\ref{B_ij_pert_almost_general_results})-(\ref{B_rphi_orthonormal_P}), so that it is straightforward to compute the eigenvalues and eigenvectors directly by diagonalization.

We find that the first eigenvalue vanishes to our order of analysis,
\begin{equation} 
\lambda_{B1}= {\mathcal O}(\epsilon_{P}^{3}),
\end{equation}
while the two other eigenvalues take the values
\begin{equation}
\lambda_{B2}= - \lambda_{B3} = -\epsilon_{P}\frac{1536{\mathcal M}r^{6}\sin\theta}{({\mathcal M}+2r)^{9}} + {\mathcal O}(\epsilon_{P}^{3}),
\end{equation} 
The corresponding orthonormal eigenvectors are
\begin{equation}
v_{B1}^{\hat \imath}
=(e_{\hat \theta})^{\hat \imath},
\end{equation}
\begin{equation}
v_{B2}^{\hat \imath}=\frac{1}{\sqrt{2}}\left((e_{\hat r})^{\hat \imath}+(e_{\hat \phi})^{\hat \imath}\right) + {\mathcal O}(\epsilon_{P}^2),
\end{equation}
and
\begin{equation}
v_{B3}^{\hat \imath}=\frac{1}{\sqrt{2}}\left((e_{\hat r})^{\hat \imath}-(e_{\hat \phi})^{\hat \imath}\right) + {\mathcal O}(\epsilon_{P}^2).
\end{equation}
Fig.~\ref{figB2z0} shows $\lambda_{B2}$ in the $z=0$ plane (orthogonal to the boost of the black hole), along with the projection of $v_{B2}^{\hat \imath}$ into that plane.  In Fig.~\ref{figB2y0} we show the same quantities, but in the $y=0$ plane.


\section{Tendex and vortex fields for binary black holes}
\label{RESULTS_BINARY}

We now turn to perturbative initial data describing an equal-mass binary black hole system in quasicircular orbit.  As demonstrated in \cite{DenBP06}, these solutions can be constructed as perturbations of two Schwarzschild black holes, by adding the perturbations created by the black holes' boosts to those created by the presence of the binary companion.  Here we follow a similar approach to construct the tendex and vortex fields for such a binary system.  While the calculations can be carried out analytically at least in principle, some of the expressions become very unwieldy and do not provide much insight.  In some cases we therefore restrict the analysis to certain regions or symmetry planes that allow a direct comparison with the results for single boosted black holes in Sec. \ref{RESULTS_SINGLEBOOSTED}.

\subsection{The Electric and Magnetic Parts of the Weyl Tensor}

The perturbations of the electric part of the Weyl tensor can be constructed as for a single boosted black hole, adding contributions from the two black holes' boosts and then taking into account the perturbation created by the binary companion.   The resulting expressions, however, are very messy, and we therefore restricted this analysis to regions either close to one of the black holes, or far from both.  We will return to this analysis later.

To compute the magnetic part of the Weyl tensor we can also add the perturbations created by the two holes.  As shown in Appendix \ref{APP_CALCS_BINARY_MAGNETIC}, it is convenient to first convert the results for a single boosted black hole to a Cartesian orthonormal basis with the origin placed at the center of mass.  In this coordinate system, the two black holes, separated by a coordinate distance $s$, are placed at coordinate locations $(\pm s/2,0,0)$ with boosts $(0,\pm P,0)$.  Keeping symmetry in mind, the nonzero components of ${\mathcal B}_{\hat{\imath}\hat{\jmath}}$, to the desired order in $P$, are
\begin{eqnarray}
{\mathcal B}_{\hat{x}\hat{x}} &=& -\frac{2z}{y}{\mathcal B}_{\hat{y}\hat{z}} = -{\mathcal B}_{\hat{z}\hat{z}} = \epsilon_{P}1536{\mathcal M}z\times\\
&&\left( \frac{r_{\rm{C_{1}}}^{4}(s-2x)}{({\mathcal M}+2r_{\rm{C_{1}}})^{9}}+\frac{r_{\rm{C_{2}}}^{4}(s+2x)}{({\mathcal M}+2r_{\rm{C_{2}}})^{9}} \right)+{\mathcal O}(\epsilon_{P}^{3})\nonumber,
\end{eqnarray}
\begin{eqnarray}
{\mathcal B}_{\hat{x}\hat{y}} &=& \epsilon_{P}1536{\mathcal M}yz \left(\frac{r_{\rm{C_{2}}}^{4}}{({\mathcal M}+2r_{\rm{C_{2}}})^{9}}-\frac{r_{\rm{C_{1}}}^{4}}{({\mathcal M}+2r_{\rm{C_{1}}})^{9}}\right)
\nonumber\\&&+\:{\mathcal O}(\epsilon_{P}^{3}),
\end{eqnarray}
and
\begin{eqnarray}
{\mathcal B}_{\hat{x}\hat{z}} &=& \epsilon_{P}384{\mathcal M}\Bigg(\frac{r_{\rm{C_{1}}}^{4}\left((s-2x)^{2}-4z^{2}\right)}{({\mathcal M}+2r_{\rm{C_{1}}})^{9}}\nonumber\\&& -\:\frac{r_{\rm{C_{2}}}^{4}\left((s+2x)^{2}-4z^{2}\right)}{({\mathcal M}+2r_{\rm{C_{2}}})^{9}}\Bigg)+{\mathcal O}(\epsilon_{P}^{3}).
\end{eqnarray}

\subsection{Horizon Tendicity and Vorticity}

To calculate the horizon vorticity it is convenient to change coordinates and consider ${\mathcal B}_{ij}$ in the vicinity of one of the holes (say hole $1$ at coordinate location $(s/2,0,0)$).  In a spherical polar orthonormal basis centered on this hole, with the boost in the (new) $\hat{z}$ direction, the leading-order components of ${\mathcal B}_{ij}$ are just those specified for the single boosted black hole in equations (\ref{B_ij_pert_almost_general_results})-(\ref{B_rphi_orthonormal_P}) -- this is because Kepler's law (\ref{virial}) implies that $1/s$ scales with $\epsilon_P^2$, so that the corrections created by the companion black hole are of higher order.  As shown in \cite{DenBP06}, corrections to the horizon location due to the companion black hole are also of higher order, so the horizon is still axisymmetric.  
As a consequence, the horizon vorticity still vanishes to our order of analysis.

To find the horizon tendicity for hole $1$, say, we first need to find expressions for the components of ${\mathcal E}_{ij}$ near hole $1$ -- this is done in Appendix \ref{APP_CALCS_BINARY_ELECTRIC}.  We expand ${\mathcal E}_{ij}$ as we did for the single boosted black hole in equation (\ref{E_ij_pert_almost_general_results}).  As before, the background ${\mathcal E}_{ij}^{(0)}$ is still given by equation (\ref{Err_Schwarzschild_orthonormal}) with $\psi$ replaced by $\psi_{(0)}$.  The perturbations  ${\mathcal E}_{ij}^{(2)}$ expressed in a coordinate basis are given by equations (\ref{E_rr_P2_binary}-\ref{E_phiphi_P2_binary}) in Appendix \ref{APP_CALCS_BINARY_ELECTRIC} -- they are very similar to those found for a single boosted black hole, but differ in some terms because of the presence of the companion black hole.

From ${\mathcal E}_{ij}$ we compute the horizon tendicity following the calculation in Appendix \ref{APP_CALCS_SINGLEBOOSTED} for single black holes, expect that we replace the conformal factor with that for a binary.  The result is
\begin{eqnarray}
{\mathcal E}_{NN} &=& -\frac{1}{4{\mathcal M}^{2}}+\epsilon_{P}^{2} \frac{9}{16{\mathcal M}^{2}}\\&&+\:\epsilon_{P}^{2} \frac{(-1871+2688\ln 2)}{640 {\mathcal M}^{2}}P_{2}(\cos\theta)+{\mathcal O}(\epsilon_{P}^{4})\nonumber.
\end{eqnarray}
As for the single boosted black hole, we can simplify the first two terms by expressing them in terms of the irreducible mass.  The relationship between the bare mass and the irreducible mass in a binary is given by equation (47) in \cite{DenBP06},
\begin{equation}
{\mathcal M} = M_{\rm irr} - \frac{P^{2}}{8M_{\rm irr}} - \frac{M_{\rm irr}^{2}}{2s} + {\mathcal O}(\epsilon_{P}^{4}).
\end{equation}
We therefore see that the average horizon tendicity is again 
\begin{equation}
{\mathcal E}_{NN}^{\rm ave} = -\frac{1}{4M_{\rm irr}^{2}} + {\mathcal O}(\epsilon_{P}^{4}).
\end{equation}
To leading order, the deviation from this average value is identical to that for a single boosted black hole, so a plot of the deviation would look like Fig. \ref{fig1}.  The spin coefficient argument made in Appendix \ref{APP_CALCS_SINGLEBOOSTED_HORIZON} for the single boosted black hole applies here as well - equations (\ref{horizon_tendicity_spincoefficients}-\ref{apply_Gauss_Bonnet_here}) are unchanged even though expressions for their constituents change slightly.

\subsection{Vortex Fields}

While the eigenvalues and eigenvectors of the magnetic part of the Weyl tensor can be computed everywhere, they are in general quite complicated and do not offer much insight.  We instead restrict our analysis to certain planes, some of which allow for a direct comparison with the results for a single boosted black hole.

\subsubsection{The symmetry plane between the black holes}

\begin{figure}
\includegraphics[width=0.48\textwidth]{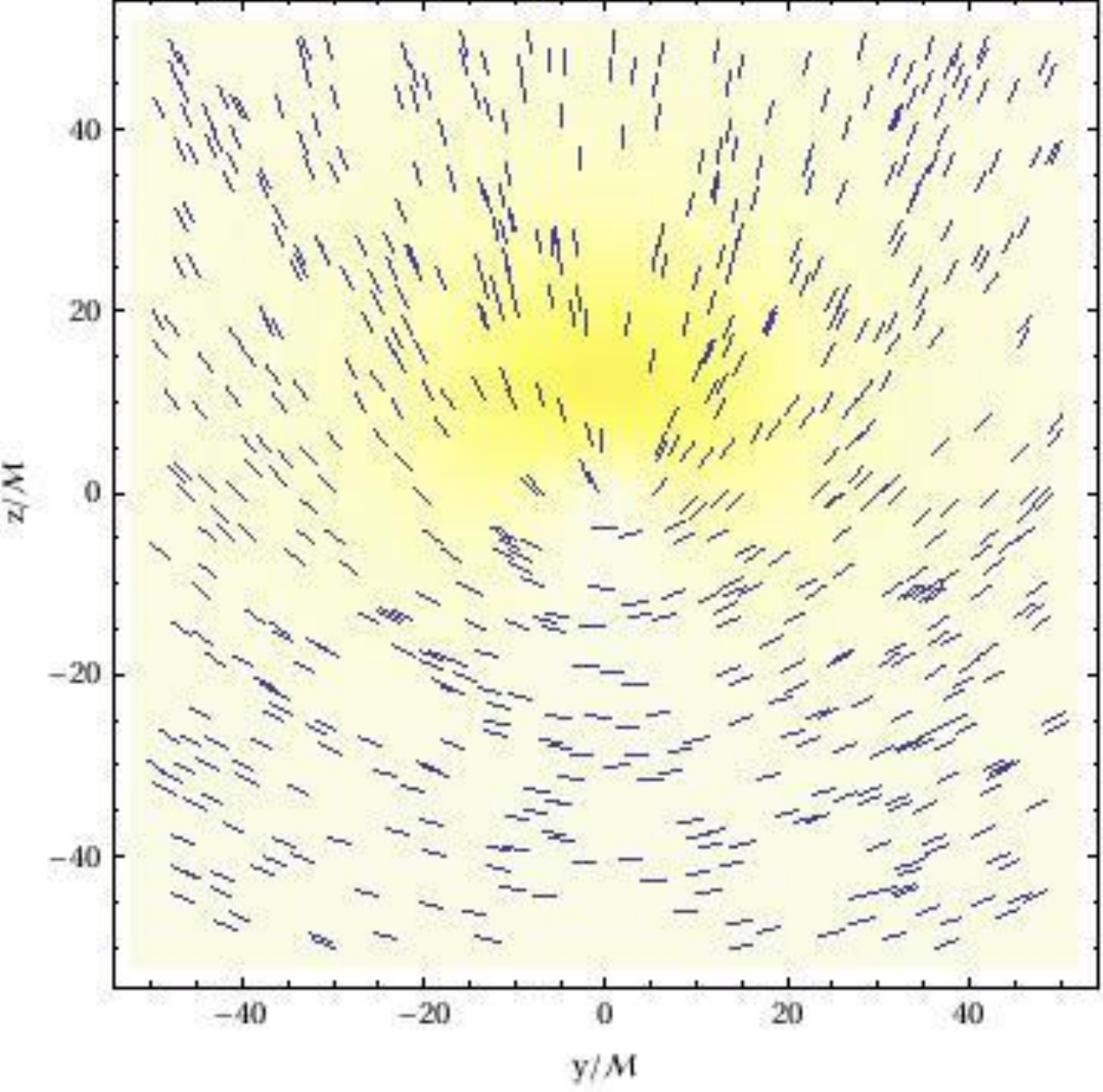}
\caption{The vortex eigenvalue $\lambda_{B2}$ in the symmetry plane $x=0$ between the two binary companions, along with the projection of eigenvector $v_{B2}^{\hat \imath}$ into that plane for a binary at separation $s = 50 {\mathcal M}$.  The analogous figure for $\lambda_{B3}$ and $v_{B3}^{\hat \imath}$ would look like this figure flipped vertically about $z=0$, but with the degree of shading indicating the magnitude of a positive eigenvalue.}
\label{figB2x0binary}
\end{figure}

The $x=0$ symmetry plane midway between the black holes provides the simplest case.   Defining $r_{C} \equiv r_{C_{1}}=r_{C_{2}}=\sqrt{(s/2)^{2} + y^{2} + z^{2}}$ we can write the eigenvalues as 
\begin{equation}
\lambda_{B1} = \epsilon_{P}\frac{3072{\mathcal M}r_{C}^{4}sz}{({\mathcal M}+2r_{C})^{9}}\left(1 + {\mathcal O}(\epsilon_{P}^{2})\right),
\end{equation} 
\begin{equation}
\lambda_{B2} = -\epsilon_{P}\frac{1536{\mathcal M}r_{C}^{4}s(z+\sqrt{y^{2} + z^{2}})}{({\mathcal M}+2r_{C})^{9}}\left(1 + {\mathcal O}(\epsilon_{P}^{2})\right),
\end{equation}
and
\begin{equation}
\lambda_{B3} = \epsilon_{P}\frac{1536{\mathcal M}r_{C}^{4}s(-z+\sqrt{y^{2} + z^{2}})}{({\mathcal M}+2r_{C})^{9}}\left(1 + {\mathcal O}(\epsilon_{P}^{2})\right).
\end{equation}
The corresponding eigenvectors are 
\begin{equation}
v_{B1}^{\hat \imath} = (1,0,0) + {\mathcal O}(\epsilon_{P}^{2}),
\end{equation}
\begin{equation}
v_{B2}^{\hat \imath} = A \bigg(0,y,z+\sqrt{y^{2} + z^{2}}\bigg) + {\mathcal O}(\epsilon_{P}^{2}),
\end{equation}
and
\begin{equation}
v_{B3}^{\hat \imath} = A \bigg(0,-(z+\sqrt{y^{2} + z^{2}}),y\bigg) + {\mathcal O}(\epsilon_{P}^{2}),
\end{equation}
where the normalization factor $A$ is given by
\begin{equation}
A = \frac{1}{\sqrt{2}\sqrt{y^{2} + z(z+\sqrt{y^2 + z^2})}}.
\end{equation}
In Fig.~\ref{figB2x0binary} we show the eigenvalue $\lambda_{B2}$ in the $x=0$ symmetry plane along with the projection of eigenvector $v_{B2}^{\hat \imath}$ into that plane.  In this figure, and in all the following figures, we assume a binary separation $s = 50 {\mathcal M}$.

\subsubsection{The symmetry plane orthogonal to the black holes' boosts}

\begin{figure}
\includegraphics[width=0.48\textwidth]{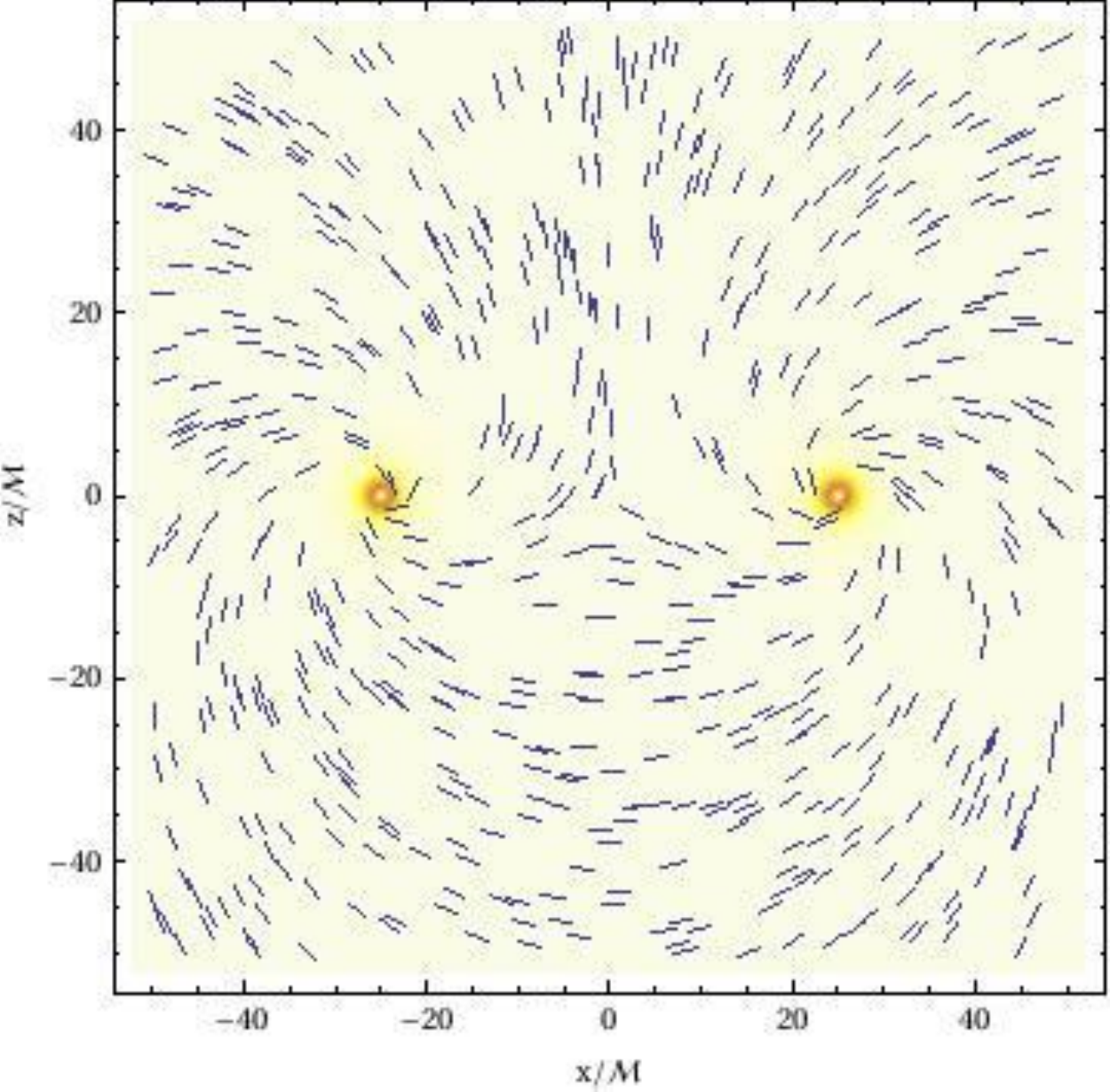}
\caption{The vortex eigenvalue $\lambda_{B2}$ in the symmetry plane $y=0$ orthogonal to the black holes' boosts, along with the projection of eigenvector $v_{B2}^{\hat \imath}$ into that plane.  A close-up view of the region near hole $1$ is shown in Fig.~\ref{figB2y0binaryclose}.}
\label{figB2y0binary}
\end{figure}

\begin{figure}
\includegraphics[width=0.48\textwidth]{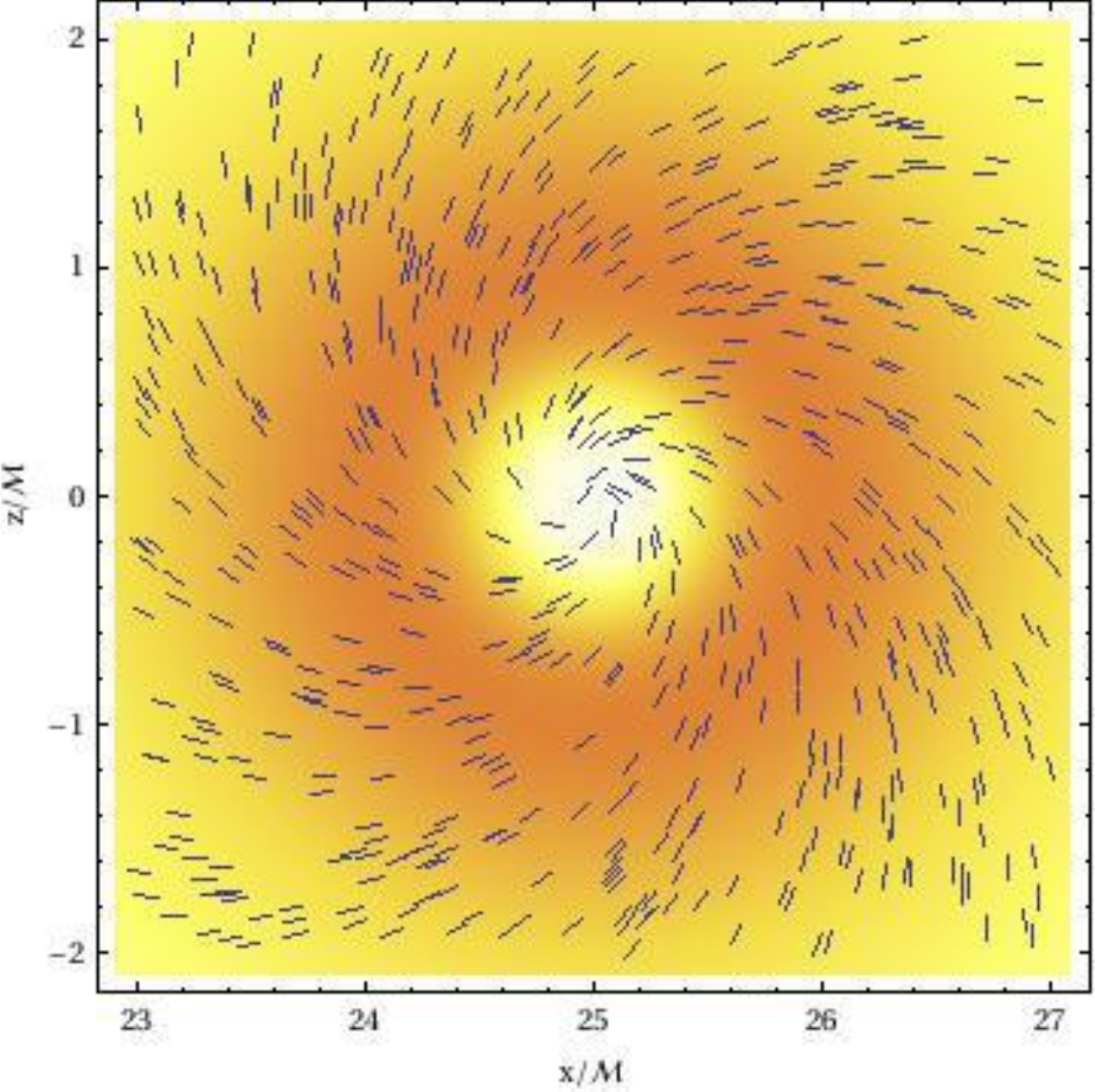}
\caption{Same as Fig.~\ref{figB2y0binary}, but in the vicinity of hole $1$.  Compare to Fig. \ref{figB2z0} - note that hole $1$ is boosted into the page here while the hole in Fig. \ref{figB2z0} is boosted out of the page.}
\label{figB2y0binaryclose}
\end{figure}  

Expressions for the eigenvalues and eigenvectors in the $y=0$ and $z=0$ planes are somewhat more complicated but can be written compactly in terms of components of ${\mathcal B}_{\hat{\imath}\hat{\jmath}}$.  For the $y=0$ symmetry plane, which is orthogonal to the black holes' boosts,  the eigenvalues are
\begin{equation}
\lambda_{B1} = {\mathcal O}(\epsilon_{P}^{3}),
\end{equation}
\begin{equation}
\lambda_{B2} = -\sqrt{({\mathcal B}_{\hat{x}\hat{x}})^{2}+({\mathcal B}_{\hat{x}\hat{z}})^{2}} + {\mathcal O}(\epsilon_{P}^{3}),
\end{equation}
and
\begin{equation}
\lambda_{B3} = \sqrt{({\mathcal B}_{\hat{x}\hat{x}})^{2}+({\mathcal B}_{\hat{x}\hat{z}})^{2}} + {\mathcal O}(\epsilon_{P}^{3}).
\end{equation}
The corresponding eigenvectors are 
\begin{equation}
v_{B1}^{\hat \imath} = (0,1,0) + {\mathcal O}(\epsilon_{P}^{2}),
\end{equation}
\begin{eqnarray}
v_{B2}^{\hat \imath} &=& A({\mathcal B}_{\hat{x}\hat{x}}-\sqrt{({\mathcal B}_{\hat{x}\hat{x}})^{2}+({\mathcal B}_{\hat{x}\hat{z}})^{2}},0,{\mathcal B}_{\hat{x}\hat{z}})\nonumber\\&&+\:{\mathcal O}(\epsilon_{P}^{2}),
\end{eqnarray}
and
\begin{eqnarray}
v_{B3}^{\hat \imath} &=& A({\mathcal B}_{\hat{x}\hat{x}}+\sqrt{({\mathcal B}_{\hat{x}\hat{x}})^{2}+({\mathcal B}_{\hat{x}\hat{z}})^{2}},0,{\mathcal B}_{\hat{x}\hat{z}})\nonumber\\&&+\:{\mathcal O}(\epsilon_{P}^{2}),
\end{eqnarray}
where the normalization factor $A$ is given by
\begin{equation}
A = \frac{1}{\sqrt{2}\sqrt{({\mathcal B}_{\hat{x}\hat{x}})^{2} + ({\mathcal B}_{\hat{x}\hat{z}})^{2} + {\mathcal B}_{\hat{x}\hat{x}}\sqrt{({\mathcal B}_{\hat{x}\hat{x}})^{2}+({\mathcal B}_{\hat{x}\hat{z}})^{2}}}}.
\end{equation}
In Fig.~\ref{figB2y0binary} we show the eigenvalue $\lambda_{B2}$ in the $y=0$  plane, together with the projection of eigenvector $v_{B2}^{\hat \imath}$ into that plane.  A close-up view of the region near hole $1$ is shown in Fig.~\ref{figB2y0binaryclose}.

\subsubsection{The orbital plane}

\begin{figure}
\includegraphics[width=0.48\textwidth]{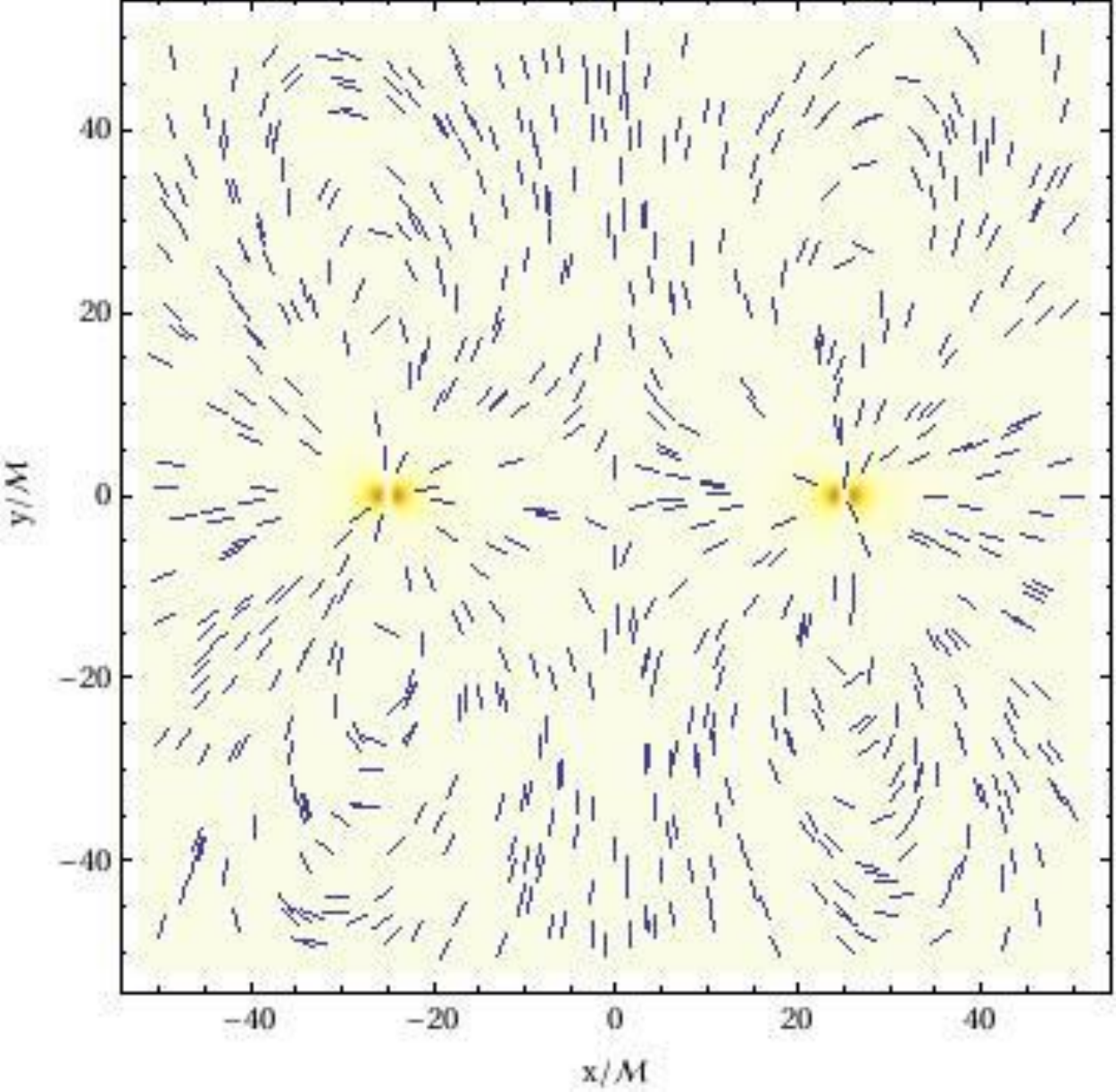}
\caption{The vortex eigenvalue $\lambda_{B2}$ in the orbital plane $z=0$ along with the projection of eigenvector $v_{B2}^{\hat \imath}$ into that plane.  A close-up view of the region near hole $1$ is shown in Fig. \ref{figB2z0binaryclose}.}
\label{figB2z0binary}
\end{figure}

\begin{figure}
\includegraphics[width=0.48\textwidth]{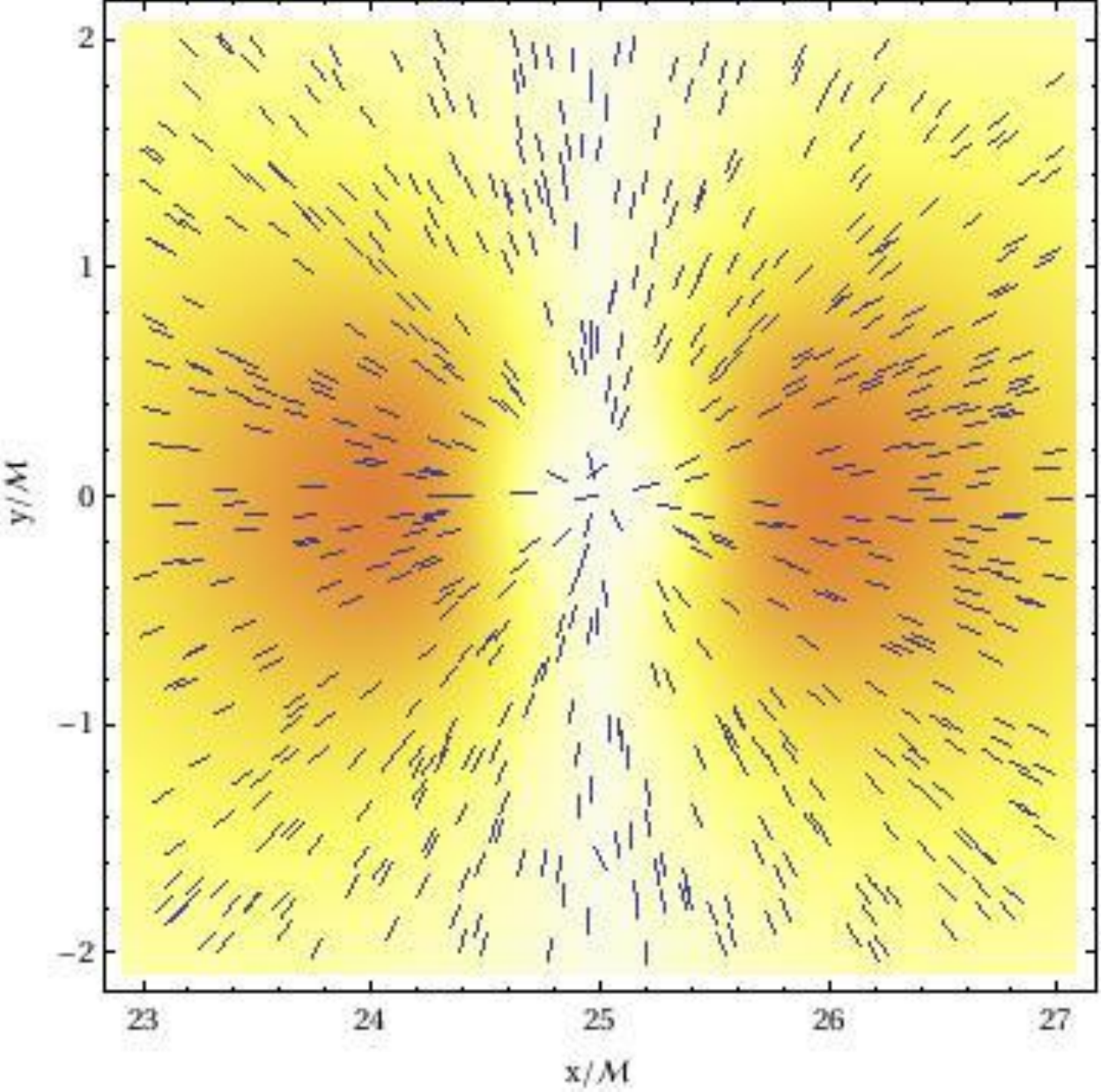}
\caption{Same as Fig.~\ref{figB2z0binary}, but in the vicinity of hole $1$.  Compare to Fig. \ref{figB2y0}.}
\label{figB2z0binaryclose}
\end{figure}

Finally, in the orbital plane $z=0$ the eigenvalues are
\begin{equation}
\lambda_{B1} = {\mathcal O}(\epsilon_{P}^{3}),
\end{equation}
\begin{equation}
\lambda_{B2} = -\sqrt{({\mathcal B}_{\hat{x}\hat{z}})^{2} + ({\mathcal B}_{\hat{y}\hat{z}})^{2}} + {\mathcal O}(\epsilon_{P}^{3}),
\end{equation}
and
\begin{equation}
\lambda_{B3} = \sqrt{({\mathcal B}_{\hat{x}\hat{z}})^{2} + ({\mathcal B}_{\hat{y}\hat{z}})^{2}} + {\mathcal O}(\epsilon_{P}^{3}).
\end{equation}
The corresponding eigenvectors are
\begin{eqnarray}
v_{B1}^{\hat \imath} &=& A\sqrt{2}(-{\mathcal B}_{\hat{y}\hat{z}},{\mathcal B}_{\hat{x}\hat{z}},0)+{\mathcal O}(\epsilon_{P}^{2}),
\end{eqnarray}
\begin{eqnarray}
v_{B2}^{\hat \imath} &=& A\left(-{\mathcal B}_{\hat{x}\hat{z}},-{\mathcal B}_{\hat{y}\hat{z}},\sqrt{({\mathcal B}_{\hat{x}\hat{z}})^{2} + ({\mathcal B}_{\hat{y}\hat{z}})^{2}}\right)\nonumber\\&&+\:{\mathcal O}(\epsilon_{P}^{2}),
\end{eqnarray}
and
\begin{eqnarray}
v_{B3}^{\hat \imath} &=& A\left({\mathcal B}_{\hat{x}\hat{z}},{\mathcal B}_{\hat{y}\hat{z}},\sqrt{({\mathcal B}_{\hat{x}\hat{z}})^{2} + ({\mathcal B}_{\hat{y}\hat{z}})^{2}}\right)\nonumber\\&&+\:{\mathcal O}(\epsilon_{P}^{2}),
\end{eqnarray}
where the normalization factor $A$ is given by
\begin{equation}
A = \frac{1}{\sqrt{2}\sqrt{({\mathcal B}_{\hat{x}\hat{z}})^{2} + ({\mathcal B}_{\hat{y}\hat{z}})^{2}}}.
\end{equation}
Fig. \ref{figB2z0binary} shows eigenvalue $\lambda_{B2}$ in the plane $z=0$ along with the projection of eigenvector $v_{B2}^{\hat \imath}$ into that plane.  A close-up view of the region near hole $1$ is shown in Fig. \ref{figB2z0binaryclose}.

\subsection{Tendex Fields}

While the electric part of the Weyl tensor can, in principle, be computed everywhere, it is in general quite complicated, meaning that the expressions for its eigenvalues and eigenvectors are even more unwieldy.  We therefore restrict our analysis to two regions, namely the vicinity of one of the black holes, and the asymptotically far region.

\subsubsection{Vicinity of one black hole}

As before, we use the perturbation theory results summarized in Appendix \ref{APP_QMGR} to find the perturbed eigenvalues and eigenvectors of ${\mathcal E}_{\hat{\imath}\hat{\jmath}}$ near hole $1$.  We change coordinates as in the horizon vorticity calculation above, replacing $r_{C_{1}}$ with $r$ for notational convenience.  To second order in $\epsilon_{P}$, the perturbed eigenvectors are exactly those given by equations 
(\ref{singleboostedvE1pieces}), (\ref{singleboostedvE2}) and (\ref{singleboostedvE3}) for the single boosted black hole, but the eigenvalues are slightly different for this problem.  The expansions (\ref{lambdaE1pieces}), (\ref{lambdaE2pieces}), and (\ref{lambdaE3pieces}) apply as before, but the perturbative pieces change slightly.  The perturbative piece of the first eigenvalue is given by
\begin{eqnarray}
\lambda_{E1}^{(2)} &=& \frac{4 {\mathcal M}}{5 ({\mathcal M}+2
   r)^{12}} \bigg(-21 {\mathcal M}^2 \Big(3 \big(\cos 2 \theta \big)+1\Big)\nonumber\\&&\times\: \Big({\mathcal M}^2+2
   {\mathcal M} r+24 r^2\Big)\Big({\mathcal M}+2 r\Big)^5\nonumber\\&&\times\:\ln \Big(\frac{{\mathcal M}}{{\mathcal M}+2 r}\Big)-6 {\mathcal M} r
   \Big(\cos 2 \theta \Big) \times\nonumber\\&&\Big(21 {\mathcal M}^7+231 {\mathcal M}^6 r+1540 {\mathcal M}^5 r^2+6930 {\mathcal M}^4
   r^3\nonumber\\&&+\:18720 {\mathcal M}^3 r^4+27568 {\mathcal M}^2 r^5+18816 {\mathcal M} r^6\nonumber\\&&+\:3360 r^7\Big)+2 r
   \Big(-21 {\mathcal M}^8-231 {\mathcal M}^7 r\nonumber\\&&-\:1630 {\mathcal M}^6 r^2-6930 {\mathcal M}^5 r^3-13320 {\mathcal M}^4 r^4\nonumber\\&&+\:1232
   {\mathcal M}^3 r^5+45984 {\mathcal M}^2 r^6+65760 {\mathcal M} r^7\nonumber\\&&+\:22400 r^8\Big)\bigg).
\end{eqnarray}
The perturbative pieces of the second and third eigenvalues are given by equations (\ref{lambdaE2pertpieceP}) and (\ref{lambdaE3pertpieceP}).  The auxiliary variables $A$ and $C$ remain the same as in the single boosted black hole case, but $B$ and $D$ are now given by
\begin{eqnarray}
B &=& 21 {\mathcal M}^8 + 294 {\mathcal M}^7
   r+1432 {\mathcal M}^6 r^2+3234 {\mathcal M}^5 r^3\nonumber\\&&+\:4908 {\mathcal M}^4 r^4+11872 {\mathcal M}^3 r^5+28320 {\mathcal M}^2
   r^6\nonumber\\&&+\:35040 {\mathcal M} r^7+11200 r^8
\end{eqnarray}
and
\begin{eqnarray}
D &=& 42 {\mathcal M}^8+525 {\mathcal M}^7 r+3062 {\mathcal M}^6 r^2 + 10164 {\mathcal M}^5
   r^3\nonumber\\&&+\:18228 {\mathcal M}^4 r^4+10640 {\mathcal M}^3 r^5 - 17664 {\mathcal M}^2 r^6\nonumber\\&&-\:30720 {\mathcal M} r^7-11200r^8.
\end{eqnarray}

\subsubsection{Asymptotic region}

We can also find the eigenvalues and eigenvectors of ${\mathcal E}_{\hat{\imath}\hat{\jmath}}$ for the binary far from both holes.  We return to our original coordinate system for the binary, with $r=\sqrt{x^{2}+y^{2}+z^{2}}$ measured from the center of mass.  For $r \gg s$, we can expand the conformal factor (\ref{pertbinpsisol}) in powers of $\epsilon_{r} \equiv s/r$ to find
\begin{equation}
\label{psi_farfield}
\psi = 1 + \frac{1}{2r} \left( 2 {\mathcal M} + 2 \frac{5 P^2}{8 {\mathcal M}} \right) + {\mathcal O}(\epsilon_{P}^{6}\epsilon_{r})
\end{equation}
(where $P$ is the magnitude of each individual black hole's momentum).   We can therefore 
identify the ADM mass of this system as 
\begin{equation}
M_{\rm ADM} =  2 \left( {\mathcal M} + \frac{5 P^2}{8 {\mathcal M}}\right) + {\mathcal O}(\epsilon_{P}^{4})
\end{equation}
(see also equation (44) in \cite{DenBP06}), where the first term on the right hand side accounts for the bare masses of the two black holes, and the second term for the binding and kinetic energies.   To leading order, therefore, the conformal factor of the binary appears like the conformal factor of a single black hole with the above ADM mass located at the origin.

To compute the electric part of the Weyl tensor, we recognize that the leading order terms in the Ricci tensor $R_{ij}$ arise from the ADM mass term in the conformal factor above, and that the contributions from the extrinsic curvature fall off more rapidly and can be neglected.  To leading order, therefore, the electric part of the Weyl tensor is identical to that of a single Schwarzschild black hole, if we express its mass in terms of the ADM mass.  Borrowing the results from Section \ref{RESULTS_SCHWARZSCHILD} we have
\begin{equation}
{\mathcal E}_{\hat{r}\hat{r}} = -2{\mathcal E}_{\hat{\theta}\hat{\theta}} = -2{\mathcal E}_{\hat{\phi}\hat{\phi}} 
= - \frac{2M_{\rm ADM}}{r^3} + {\mathcal O}(\epsilon_{P}^{6}\epsilon_{r}^3).
\end{equation}
Similarly, the tendex fields for the binary in the asymptotic region are identical to those for a single black hole derived in Section \ref{RESULTS_SCHWARZSCHILD}, just with the bare mass replaced by the ADM mass.


\section{Summary}
\label{SUM}

The authors of \cite{OweBCKLMNSZZT11} recently introduced tendex and vortex fields, defined in terms of the eigenvectors and eigenvalues of the electric and magnetic parts of the Weyl tensor, as an aid to visualize spacetime curvature.   In particular, the method has promise to help interpret results from numerical relativity simulations, and to provide insight into the physical processes governing the coalescence and merger of binary black holes as well as the emission of gravitational radiation.

Many numerical simulations of binary black holes start with wormhole initial data constructed with the puncture method (see, e.g., \cite{BraB97,Bau00}).  Here we present perturbative but analytical expressions for the tendex and vortex fields, based on the perturbative treatment of these initial data presented in \cite{DenBP06}.  In this approach, the boosts of the individual black holes and the effect of the binary companion are treated as perturbations of a Schwarzschild black hole, so that the data, and hence our expressions for the tendex and vortex fields, become exact in the limit of vanishing boost or large binary separation.  

Our results complement other examples of tendex and vortex fields presented in \cite{OweBCKLMNSZZT11,NicOZZBCKLMST11,ZimNZ11}, and help us to better understand the properties of these fields.   We hope that, as analytical expressions for strong-field objects, our results may also be useful for comparison with future numerical calculations.


\acknowledgments

It is a pleasure to thank Steve Naculich for useful conversations, and David Brown for creating GRwiki, which has proven helpful for several aspects of this paper.  This work was supported in part by NSF Grant PHY-1063240 to Bowdoin College.


\begin{appendix}

\section{Finding Eigenvalues and Eigenvectors Perturbatively}
\label{APP_QMGR}

Finding the eigenvalues and eigenvectors of a perturbed matrix is a common problem in quantum mechanics that in its simplest form can be handled by stationary perturbation theory.  It is less familiar in this context, so in this appendix we review well-known results (see, e.g., \cite{CohDL77,Bay90}) just to the extent needed for the calculations in Appendix \ref{APP_CALCS_SINGLEBOOSTED} and Appendix \ref{APP_CALCS_BINARY}.  Note that this is first order perturbation theory, but that for consistency with the application and notation elsewhere in the paper, we work to first order in a small quantity called $\epsilon^{2}$.  We also choose to work in an orthonormal basis.  

Let ${\mathcal E}_{\hat{\imath}\hat{\jmath}}^{(0)}$ be a $3\times3$ symmetric matrix with real entries.  ${\mathcal E}_{\hat{\imath}\hat{\jmath}}^{(0)}$ has three real eigenvalues $\lambda_{E1}^{(0)}$, $\lambda_{E2}^{(0)}$, and $\lambda_{E3}^{(0)}$, and corresponding orthonormal eigenvectors $v^{(0)\:\hat{\imath}}_{E1}$, $v^{(0)\:\hat{\imath}}_{E2}$, and $v^{(0)\:\hat{\imath}}_{E3}$.  We now consider a perturbation of ${\mathcal E}_{\hat{\imath}\hat{\jmath}}^{(0)}$ of the form
\begin{equation}
\label{perturbed_matrix}
{\mathcal E}_{\hat{\imath}\hat{\jmath}} = {\mathcal E}_{\hat{\imath}\hat{\jmath}}^{(0)} + \epsilon^{2}{\mathcal E}_{\hat{\imath}\hat{\jmath}}^{(2)}.
\end{equation}
We then want to find the new eigenvalues
\begin{eqnarray}
\label{perturbed_eigenvalues}
\lambda_{E1} &=& \lambda_{E1}^{(0)} + \epsilon^{2}\lambda_{E1}^{(2)} + {\mathcal O}(\epsilon^{4}),\nonumber\\
\lambda_{E2} &=& \lambda_{E2}^{(0)} + \epsilon^{2}\lambda_{E2}^{(2)} + {\mathcal O}(\epsilon^{4}),\:\rm{and}\nonumber\\
\lambda_{E3} &=& \lambda_{E3}^{(0)} + \epsilon^{2}\lambda_{E3}^{(2)} + {\mathcal O}(\epsilon^{4}), 
\end{eqnarray}
along with corresponding new eigenvectors normalized to order $\epsilon^{2}$
\begin{eqnarray}
\label{perturbed_eigenvectors}
v^{\hat{\imath}}_{E1} &=& v^{(0)\:\hat{\imath}}_{E1} + \epsilon^{2}v^{(2)\:\hat{\imath}}_{E1} + {\mathcal O}(\epsilon^{4}),\nonumber\\
v^{\hat{\imath}}_{E2} &=& v^{(0)\:\hat{\imath}}_{E2} + \epsilon^{2}v^{(2)\:\hat{\imath}}_{E2} + {\mathcal O}(\epsilon^{4}),\:\rm{and}\nonumber\\
v^{\hat{\imath}}_{E3} &=& v^{(0)\:\hat{\imath}}_{E3} + \epsilon^{2}v^{(2)\:\hat{\imath}}_{E3} + {\mathcal O}(\epsilon^{4}).
\end{eqnarray}
Exactly how we do this depends upon the degeneracy of the eigenvalues.


\subsection{Nondegenerate Eigenvalues}
\label{APP_QMGR_NONDEG}

Suppose that $\lambda_{E1}^{(0)}$ is a nondegenerate eigenvalue.  The corresponding perturbed eigenvalue is then
\begin{equation}
\label{perturbed_eigenvalue_nondeg}
\lambda_{E1} = \lambda_{E1}^{(0)} + \epsilon^{2} {\mathcal E}_{ij}^{(2)}v^{(0)\:\hat{\imath}}_{E1}v^{(0)\:\hat{\jmath}}_{E1} + {\mathcal O}(\epsilon^{4}),
\end{equation}
and the corresponding perturbed eigenvector is
\begin{eqnarray}
\label{perturbed_eigenvector_nondeg}
v^{\hat{\imath}}_{E1} &=& v^{(0)\:\hat{\imath}}_{E1} + \epsilon^{2}\frac{{\mathcal E}_{\hat{\jmath}\hat{k}}^{(2)}v^{(0)\:\hat{\jmath}}_{E1}v^{(0)\:\hat{k}}_{E2}}{\lambda_{E1}^{(0)}-\lambda_{E2}^{(0)}}v^{(0)\:\hat{\imath}}_{E2} \nonumber\\&&+\: 
\epsilon^{2}\frac{{\mathcal E}_{\hat{\jmath}\hat{k}}^{(2)}v^{(0)\:\hat{\jmath}}_{E1}v^{(0)\:\hat{k}}_{E3}}{\lambda_{E1}^{(0)}-\lambda_{E3}^{(0)}}v^{(0)\:\hat{\imath}}_{E3} + {\mathcal O}(\epsilon^{4}).
\end{eqnarray}


\subsection{Degenerate Eigenvalues}
\label{APP_QMGR_DEG}

Suppose that the unperturbed matrix ${\mathcal E}_{\hat{\imath}\hat{\jmath}}^{(0)}$ has a two-fold degeneracy, e.g. $\lambda_{E2}^{(0)}=\lambda_{E3}^{(0)}$.  In this case the normalized vectors $v^{(0)\:\hat{\imath}}_{E2}$ and $v^{(0)\:\hat{\imath}}_{E3}$ might be any two orthonormal vectors spanning the corresponding subspace.  Choose $v^{(0)\:\hat{\imath}}_{E2}$ and $v^{(0)\:\hat{\imath}}_{E3}$ so that $\epsilon^{2}{\mathcal E}_{\hat{\imath}\hat{\jmath}}^{(2)}v^{(0)\:\hat{\imath}}_{E2}v^{(0)\:\hat{\jmath}}_{E3} = 0$.  This ensures
that the unperturbed states are indeed the limit of the perturbed states as $\epsilon^{2}\to 0$.  Then, the perturbed eigenvalues are
\begin{equation}
\label{perturbed_eigenvalue_v_deg}
\lambda_{E2} = \lambda_{E2}^{(0)} + \epsilon^{2}{\mathcal E}_{\hat{\imath}\hat{\jmath}}^{(2)}v^{(0)\:\hat{\imath}}_{E2}v^{(0)\:\hat{\jmath}}_{E2} + {\mathcal O}(\epsilon^{4}), 
\end{equation}
and
\begin{equation}
\label{perturbed_eigenvalue_w_deg}
\lambda_{E3} = \lambda_{E3}^{(0)} + \epsilon^{2}{\mathcal E}_{\hat{\imath}\hat{\jmath}}^{(2)}v^{(0)\:\hat{\imath}}_{E3}v^{(0)\:\hat{\jmath}}_{E3} + {\mathcal O}(\epsilon^{4}), 
\end{equation}
and if the perturbation succeeds in breaking the degeneracy at first order in $\epsilon^{2}$, the eigenvectors are
\begin{equation}
\label{perturbed_eigenvector_v_deg}
v^{\hat{\imath}}_{E2}  = v^{(0)\:\hat{\imath}}_{E2}  + \epsilon^{2}\frac{{\mathcal E}_{\hat{\jmath}\hat{k}}^{(2)}v^{(0)\:\hat{\jmath}}_{E2}v^{(0)\:\hat{k}}_{E1}}{\lambda_{E2}^{(0)}-\lambda_{E1}^{(0)} }v^{(0)\:\hat{\imath}}_{E1} + 
{\mathcal O}(\epsilon^{4}),
\end{equation}
and
\begin{equation}
\label{perturbed_eigenvector_w_deg}
v^{\hat{\imath}}_{E3}  = v^{(0)\:\hat{\imath}}_{E3}  + \epsilon^{2}\frac{{\mathcal E}_{\hat{\jmath}\hat{k}}^{(2)}v^{(0)\:\hat{\jmath}}_{E3}v^{(0)\:\hat{k}}_{E1}}{\lambda_{E3}^{(0)}-\lambda_{E1}^{(0)} }v^{(0)\:\hat{\imath}}_{E1} + 
{\mathcal O}(\epsilon^{4}).
\end{equation}


\section{Single Boosted Black Holes}
\label{APP_CALCS_SINGLEBOOSTED}

\subsection{The Electric Part of the Weyl Tensor}
\label{APP_CALCS_SINGLEBOOSTED_ELECTRIC}

From the form of the initial data specified in Sec. \ref{ID_SINGLEBOOSTED} we can expect the lowest-order perturbation to ${\mathcal E}_{ij}$ to be of order $\epsilon_{P}^{2}$:
\begin{equation}
\label{E_ij_pert_almost_general}
{\mathcal E}_{ij} = {\mathcal E}_{ij}^{(0)} + \epsilon_{P}^{2}{\mathcal E}_{ij}^{(2)}  +
{\mathcal O}(\epsilon_{P}^{4}).
\end{equation}
If we write the Ricci tensor as
\begin{equation}
R_{ij} = R_{ij}^{(0)} + \epsilon_{P}^{2}R_{ij}^{(2)} + {\mathcal O}(\epsilon_{P}^{4}),
\end{equation}
we see that
\begin{equation}
\label{E_ij_0_is_R_ij_0}
{\mathcal E}_{ij}^{(0)} = R_{ij}^{(0)},
\end{equation}
which we already computed in Sec. \ref{RESULTS_SCHWARZSCHILD}, and
\begin{equation}
\label{E_ij_P2_Defined}
{\mathcal E}_{ij}^{(2)} = R_{ij}^{(2)} - \gamma^{kl}_{(0)}K_{il}^{(1)}K_{jk}^{(1)}.
\end{equation} 
The perturbative term ${\mathcal E}_{ij}^{(2)}$ can then be found by substituting equation (\ref{conformal_single_boosted_short}) into equations (\ref{spatial_metric_conformal_form}), (\ref{Kij_in_terms_of_Aij}), and (\ref{R_ij_in_conformal_terms}).  We note that
\begin{equation}
\label{gamma_expanded}
\gamma^{ij}_{(0)} = \psi_{(0)}^{-4}\bar{\gamma}^{ij},
\end{equation}
\begin{equation}
\label{extcurv_lowest}
K_{ij}^{(1)} = \psi_{(0)}^{-2}\bar{A}_{ij},
\end{equation}
and we compute the perturbation of the Ricci tensor from
\begin{eqnarray}
\label{R_ij_P2_Defined}
R_{ij}^{(2)} &=& -2\Big(\bar{D}_{i}\bar{D}_{j}(\psi^{-1}_{(0)}u)\nonumber\\&&+\:\bar{\gamma}_{ij}^{(0)}\bar{\gamma}^{lm}_{(0)}\bar{D}_{l}\bar{D}_{m}(\psi_{(0)}^{-1}u)\Big)\nonumber\\&&+\:4\Big(\bar{D}_{i}(\ln\psi_{(0)})\bar{D}_{j}(\psi^{-1}_{(0)}u)\nonumber\\&&+\:\bar{D}_{j}(\ln\psi_{(0)})\bar{D}_{i}(\psi^{-1}_{(0)}u)\nonumber\\&&-\:\bar{\gamma}_{ij}^{(0)}\bar{\gamma}^{lm}_{(0)}\big(\bar{D}_{l}(\ln\psi_{(0)})\bar{D}_{m}(\psi^{-1}_{(0)}u)\nonumber\\&&+\:\bar{D}_{m}(\ln\psi_{(0)})\bar{D}_{l}(\psi^{-1}_{(0)}u)\big)\Big).
\end{eqnarray} 
From (\ref{pertboostusol}) and (\ref{Aijup}) we then find the nonzero components of ${\mathcal E}_{ij}^{(2)}$ to be
\begin{eqnarray}
\label{E_rr_P2}
{\mathcal E}_{rr}^{(2)} &=& \frac{{\mathcal M}}{20 r^4 ({\mathcal M}+2 r)^8} \Bigg(3 {\mathcal M} \bigg(-2 r (\cos 2 \theta ) \Big(21 {\mathcal M}^7\nonumber\\&&+\:399 {\mathcal M}^6 r+3052
   {\mathcal M}^5 r^2+12194 {\mathcal M}^4 r^3\nonumber\\&&+\:27344 {\mathcal M}^3 r^4+33712 {\mathcal M}^2 r^5+19776 {\mathcal M} r^6\nonumber\\&&+\:3360 r^7\Big)-7 {\mathcal M} \Big(3
   (\cos 2 \theta )+1\Big) \Big({\mathcal M}+4 r\Big) \nonumber\\&&\times \Big({\mathcal M}+6 r\Big) \Big({\mathcal M}+2 r\Big)^5 \ln \Big(\frac{{\mathcal M}}{{\mathcal M}+2
   r}\Big)\bigg)\nonumber\\&&-\:2 r \bigg(21 {\mathcal M}^8+399 {\mathcal M}^7 r+3062 {\mathcal M}^6 r^2\nonumber\\&&+\:12274 {\mathcal M}^5 r^3+27544 {\mathcal M}^4
   r^4+33712 {\mathcal M}^3 r^5\nonumber\\&&+\:18976 {\mathcal M}^2 r^6+2080 {\mathcal M} r^7+3200 r^8\bigg)\Bigg),
\end{eqnarray}
\begin{eqnarray}
\label{E_rtheta_P2}
{\mathcal E}_{r\theta}^{(2)} &=& {\mathcal E}_{\theta r}^{(2)} = \frac{3 {\mathcal M}^2}{20 r^3 ({\mathcal M}+2 r)^8}\Bigg(3840 r^8 \sin \theta  \cos \theta \nonumber\\&&-(\sin 2 \theta )
   \bigg({\mathcal M}+2 r\bigg) \bigg(2 r \Big(21 {\mathcal M}^6+357 {\mathcal M}^5 r\nonumber\\&&+\:2170 {\mathcal M}^4 r^2+6342 {\mathcal M}^3 r^3+9388 {\mathcal M}^2 r^4\nonumber\\&&+\:6216
   {\mathcal M} r^5+720 r^6\Big)+21 {\mathcal M} \Big({\mathcal M}+8 r\Big) \nonumber\\&&\times\Big({\mathcal M}+2 r\Big)^5\ln \Big(\frac{{\mathcal M}}{{\mathcal M}+2
   r}\Big)\bigg)\Bigg),
\end{eqnarray}
\begin{eqnarray}
\label{E_thetatheta_P2}
{\mathcal E}_{\theta\theta}^{(2)} &=& \frac{{\mathcal M}}{20 r^2 ({\mathcal M}+2 r)^8}\Bigg(3 {\mathcal M} \bigg(7 {\mathcal M} \Big({\mathcal M}+2 r\Big)^5 \nonumber\\&&\times\ln \Big(\frac{{\mathcal M}}{{\mathcal M}+2 r}\Big)
   \Big(3 (\cos 2 \theta ) \big({\mathcal M}^2+7 {\mathcal M} r\nonumber\\&&+\:14 r^2\big)-\big({\mathcal M}-2 r\big) \big({\mathcal M}+3 r\big)\Big)\nonumber\\&&+\:2 r
   (\cos 2 \theta )\Big( 21 {\mathcal M}^7+336 {\mathcal M}^6 r+2275 {\mathcal M}^5 r^2\nonumber\\&&+\:8330 {\mathcal M}^4 r^3+17528 {\mathcal M}^3
   r^4+20608 {\mathcal M}^2 r^5\nonumber\\&&+\:11664 {\mathcal M} r^6 +2400 r^7\Big)\bigg)\nonumber\\&&-\:2 r \bigg({\mathcal M}-2 r\bigg) \bigg(21 {\mathcal M}^7+252
   {\mathcal M}^6 r\nonumber\\&&+\:1220 {\mathcal M}^5 r^2+3002 {\mathcal M}^4 r^3+3800 {\mathcal M}^3 r^4\nonumber\\&&+\:2000 {\mathcal M}^2 r^5-160 {\mathcal M} r^6+800
   r^7\bigg)\Bigg),
\end{eqnarray}
and
\begin{eqnarray}
\label{E_phiphi_P2}
{\mathcal E}_{\phi\phi}^{(2)} &=& \frac{{\mathcal M}(\sin ^2\theta )}{20 r^2 ({\mathcal M}+2 r)^8} \Bigg(3 {\mathcal M} \bigg(7 {\mathcal M} \Big({\mathcal M}+2 r\Big)^5 \times\nonumber\\&&\ln
   \Big(\frac{{\mathcal M}}{{\mathcal M}+2 r}\Big) \Big(2 {\mathcal M}^2+3 r (\cos 2 \theta ) \big(3 {\mathcal M}+10 r\big)\nonumber\\&&+\:11 {\mathcal M} r+18
   r^2\Big)+6 r^2 (\cos 2 \theta ) \Big(21 {\mathcal M}^6\nonumber\\&&+\:259 {\mathcal M}^5 r+1288 {\mathcal M}^4 r^2+3272 {\mathcal M}^3
   r^3\nonumber\\&&+\:4368 {\mathcal M}^2 r^4+2704 {\mathcal M} r^5+320 r^6 \Big)\bigg)\nonumber\\&&+\:2 r \bigg(42 {\mathcal M}^8+609 {\mathcal M}^7 r+3778
   {\mathcal M}^6 r^2\nonumber\\&&+\:12836 {\mathcal M}^5 r^3+25340 {\mathcal M}^4 r^4+28112 {\mathcal M}^3 r^5\nonumber\\&&+\:14816 {\mathcal M}^2 r^6+3200 {\mathcal M} r^7+1600
   r^8 \bigg)\Bigg).
\end{eqnarray}

\subsection{The Magnetic Part of the Weyl Tensor}
\label{APP_CALCS_SINGLEBOOSTED_MAGNETIC}

The magnetic part of the Weyl tensor ${\mathcal B}_{ij}$ is defined in
equation (\ref{B_ij_Defined}).  To lowest order in the momentum this is
\begin{equation}
\label{B_ij_Pert}
{\mathcal B}_{ij} = {\mathcal B}_{ij}^{(1)} + {\mathcal O}(\epsilon_{P}^{3}) = \epsilon_{j(0)}^{lk}D_{k}^{(0)}K_{li}^{(1)} + {\mathcal O}(\epsilon_{P}^{3}), 
\end{equation}
where the extrinsic curvature is given by equation (\ref{extcurv_lowest}) and the covariant derivative and antisymmetric tensor are constructed using the unperturbed spatial metric.  Evaluating equation (\ref{B_ij_Pert}) results in 
\begin{equation}
\label{B_rphi_singleboosted_app}
{\mathcal B}_{r\phi} = {\mathcal B}_{\phi r} = -\epsilon_{P} \frac{96{\mathcal M}r^{3}\sin^{2}\theta}{({\mathcal M}+2r)^{5}} + {\mathcal O}(\epsilon_{P}^{3}),
\end{equation}
and the other components vanish to this order.

\subsection{Horizon Tendicity}
\label{APP_CALCS_SINGLEBOOSTED_HORIZON}

For the perturbative initial data of Sec.~\ref{ID_SINGLEBOOSTED}, the apparent horizon is located at a coordinate distance
\begin{equation}
\label{boosted_horizon_location}
h = \frac{{\mathcal M}}{2}-\epsilon_{P}\frac{{\mathcal M}}{16}\cos\theta+{\mathcal O}(\epsilon_{P}^{2})
\end{equation}
from the center  (see equation (24) in \cite{DenBP06}).  The inward unit normal on the horizon is then given by
\begin{equation}
\label{boosted_horizon_normal}
N^{i} = -s^{i} = -\gamma^{ij}s_{j},
\end{equation}
where 
\begin{equation}
\label{boosted_horizon_normal_s}
s_{j} = \lambda(1,-\partial_{\theta}h,0)
\end{equation}
and where the normalization factor $\lambda$ is
\begin{equation}
\label{boosted_horizon_normalization_factor}
\lambda = \left( \frac{\gamma_{rr}\gamma_{\theta\theta}-(\gamma_{r\theta})^{2}}{\gamma_{rr}(\partial_{\theta}h)^{2}+2\gamma_{r\theta}\partial_{\theta}h+\gamma_{\theta\theta}}\right)^{1/2}
\end{equation}
(see, e.g., \cite{BauS10}).  We then compute the horizon tendicity from 
\begin{equation}
\label{boosted_horizon_tendicity_1}
{\mathcal E}_{NN}={\mathcal E}_{ij}N^{i}N^{j}={\mathcal E}_{ij}\gamma^{iA}\gamma^{jB}s_{A}s_{B},
\end{equation}
where $A$ and $B$ run over only $r$ and $\theta$.  For a diagonal spatial metric this simplifies to
\begin{equation}
\label{boosted_horizon_tendicity_2}
{\mathcal E}_{NN} = {\mathcal E}_{AB}\gamma^{AA}\gamma^{BB}s_{A}s_{B}.
\end{equation}
This can be expanded in terms of unperturbed and perturbed quantities:
\begin{eqnarray}
\label{boosted_horizon_tendicity_2_exp}
{\mathcal E}_{NN} &=& R_{rr}^{(0)}(\gamma^{rr}_{(0)}+\gamma^{rr}_{(2)}-(\partial_{\theta}h)^{2}/\gamma_{\theta\theta}^{(0)})\nonumber\\&&-\:2R_{r\theta}^{(0)}\gamma_{(0)}^{\theta\theta}\partial_{\theta}h+R_{\theta\theta}^{(0)}(\gamma^{\theta\theta}_{(0)})^{2}\gamma_{rr}^{(0)}(\partial_{\theta}h)^{2}\nonumber\\&&+\:\gamma_{(0)}^{rr}R_{rr}^{(2)}-\gamma^{rr}_{(0)}\gamma_{(0)}^{kl}K_{rl}^{(1)}K_{rk}^{(1)}\nonumber\\&&+\:{\mathcal O}(\epsilon_{P}^{4}).
\end{eqnarray}
When evaluated at the horizon location, equation (\ref{boosted_horizon_tendicity_2_exp}) becomes
\begin{eqnarray}
\label{single_boosted_horizon_tendicity_app}
{\mathcal E}_{NN} &=& -\frac{1}{4{\mathcal M}^{2}} + \frac{\epsilon_{P}^{2}}{2560{\mathcal M}^{2}}\big(-1711+2688\ln 2 + 
\nonumber\\&& 3(\cos 2\theta)(-1871+2688\ln 2)\big) +{\mathcal O}(\epsilon_{P}^{4}).
\end{eqnarray}
We could plot this function for ${\mathcal E}_{NN}$, but plotting the deviation from the proper-area weighted average value of ${\mathcal E}_{NN}$ over the horizon is more interesting.  This average is 
\begin{equation}
\label{single_boosted_horizon_tendicity_avg}
{\mathcal E}_{NN}^{\rm{ave}} = \frac{\int_{0}^{2\pi}\int_{0}^{\pi} E_{NN}\psi^{4}h^{2}(1+(\partial_{\theta}h)^{2}h^{-2})^{1/2}\sin\theta d\theta d\phi}{\int_{0}^{2\pi}\int_{0}^{\pi} \psi^{4}h^{2}(1+(\partial_{\theta}h)^{2}h^{-2})^{1/2}\sin\theta d\theta d\phi},
\end{equation}
which evaluates to
\begin{equation}
\label{single_boosted_horizon_tendicity_avg_evaluated}
{\mathcal E}_{NN}^{\rm{ave}} = -\frac{1}{4{\mathcal M}^{2}}+\frac{\epsilon_{P}^{2}}{16{\mathcal M}^{2}}+{\mathcal O}(\epsilon_{P}^{4}) = -\frac{1}{4M_{\rm irr}^{2}} + {\mathcal O}(\epsilon_{P}^{4}),
\end{equation}
as discussed in Sec.~\ref{RESULTS_SINGLEBOOSTED_HORIZON}.  

The appearance of the irreducible mass in equation (\ref{single_boosted_horizon_tendicity_avg_evaluated}) is not surprising given that the denominator of equation (\ref{single_boosted_horizon_tendicity_avg}) is $16\pi M_{\rm irr}^{2} + {\mathcal O}(\epsilon_{P}^{4})$, which follows from \cite{Chr70} if we make the usual approximation of replacing the proper area of the event horizon with that of the apparent horizon.  To see that the the numerator of equation (\ref{single_boosted_horizon_tendicity_avg}) is as simple as $-4\pi + {\mathcal O}(\epsilon_{P}^{4})$, we use the Newman-Penrose formalism \cite{NewP62,PenR84} as discussed in \cite{OweBCKLMNSZZT11} to rewrite the horizon tendicity as
\begin{equation}
\label{horizon_tendicity_spincoefficients}
{\mathcal E}_{NN} = - \frac{^{(2)}\mathcal R}{2} + 2\Re(\mu\rho-\lambda\sigma),
\end{equation}
where $^{(2)}{\mathcal R}$ is the two-dimensional Ricci scalar for the horizon, $\Re$ means ``the real part of'' and $\mu$, $\rho$, $\lambda$, and $\sigma$ are spin coefficients evaluated on the horizon.  Equation (\ref{horizon_tendicity_spincoefficients}) is defined with respect to a particular null tetrad specified in \cite{OweBCKLMNSZZT11}, but the required spin coefficients can still be constructed from purely spatial quantities.  We find that
\begin{equation}
\mu = -\frac{1}{\sqrt{2}}m^{\star\:i}m^{j}(K_{ij}-D_{j}N_{i}),
\end{equation}
\begin{equation}
\rho = \frac{1}{\sqrt{2}}m^{i}m^{\star\:j}(K_{ij}+D_{j}N_{i}),
\end{equation}
\begin{equation}
\lambda = -\frac{1}{\sqrt{2}}m^{\star\:i}m^{\star\:j}(K_{ij}-D_{j}N_{i}),
\end{equation}
and
\begin{equation}
\sigma = \frac{1}{\sqrt{2}}m^{i}m^{j}(K_{ij}+D_{j}N_{i}),
\end{equation}
where $m^{j} = (e_{2}^{j}+ i e_{3}^{j})/\sqrt{2}$ is a leg of the null tetrad constructed from two orthonormal spatial vectors tangent to the horizon, and $m^{\star\:j}$ is its complex conjugate.  With the above expressions, the combination $\mu\rho-\lambda\sigma$ turns out to be purely real.  Moreover, evaluating the spin coefficients for a horizon location of the form $h={\mathcal M}/2 + \epsilon f(\theta)$ we obtain
\begin{eqnarray}
\mu\rho-\lambda\sigma &=& \frac{\epsilon_{P}^{2}}{2048{\mathcal M}^{4}}\bigg(-9{\mathcal M}^{2}\Big(\cos^{2}\theta\Big)\nonumber\\&&+\:256\Big(f-2(\cot\theta)\partial_{\theta}f\Big)\Big(f-2\partial_{\theta}\partial_{\theta}f\Big)\Bigg)\nonumber\\&&+\:{\mathcal O}(\epsilon_{P}^{4}).
\end{eqnarray}
Inserting $f(\theta)=-({\mathcal M}/16)\cos\theta$ from (\ref{boosted_horizon_location}) all the leading order terms cancel, meaning that, to our order of analysis, the integral in the numerator of (\ref{single_boosted_horizon_tendicity_avg}) reduces to the Ricci scalar term
\begin{equation}
\label{apply_Gauss_Bonnet_here}
-\frac{1}{2}\int_{0}^{2\pi}\int_{0}^{\pi} {}^{(2)}{\mathcal R}\psi^{4}h^{2}(1+(\partial_{\theta}h)^{2}h^{-2})^{1/2}\sin\theta d\theta d\phi 
\end{equation}
Since the horizon has the topology of a sphere, the Gauss-Bonnet theorem guarantees that the integral is $8\pi$, so the numerator of equation (\ref{single_boosted_horizon_tendicity_avg}) is indeed
 $-4\pi + {\mathcal O}(\epsilon_{P}^{4})$.


\section{Binary Black Holes}
\label{APP_CALCS_BINARY}

\subsection{Additivity of the Magnetic Part of the Weyl Tensor}
\label{APP_CALCS_BINARY_MAGNETIC}

We begin by showing that to calculate ${\mathcal B}_{ij}$ for the binary, we can, to our order of analysis, add the corresponding single-hole corrections, i.e.
\begin{equation}
\label{Bijaddsapp}
{\mathcal B}_{ij} = {\mathcal B}_{ij}^{C_{1}P_{1}} + {\mathcal B}_{ij}^{C_{2}P_{2}} + {\mathcal O}(\epsilon_{P}^{3}).
\end{equation}  
The conformal factor is given by equation (\ref{pertbinpsisol}),
so the extrinsic curvature is
\begin{eqnarray}
K_{ij} &=& \left(1 + \frac{{\mathcal M}_{1}}{2r_{C_{1}}} + \frac{{\mathcal M}_{2}}{2r_{C_{2}}}\right)^{-2}\left(\bar{A}_{ij}^{C_{1}P_{1}}+\bar{A}_{ij}^{C_{2}P_{2}}\right)+ \nonumber\\&& {\mathcal O}(\epsilon_{P}^{3}).
\end{eqnarray}
At a distance of $s/2$ or more away from hole $2$ we can use equation (\ref{virial}) to expand and find
\begin{equation}
\label{app_Kij_1}
K_{ij} = \left(1+\frac{{\mathcal M}_{1}}{2r_{C_{1}}}\right)^{-2}\bar{A}_{ij}^{C_{1}P_{1}}+{\mathcal O}(\epsilon_{P}^{3}).
\end{equation}
At these locations $K_{ij}^{C_{2}P_{2}} = 0 + {\mathcal O}(\epsilon_{P}^{5})$, and since we can do the analogous analysis at a distance of $s/2$ or more away from hole $1$,
we can write the extrinsic curvature everywhere as
\begin{equation}
\label{app_Kij_adds}
K_{ij} = K_{ij}^{C_{1}P_{1}}+K_{ij}^{C_{2}P_{2}}+{\mathcal O}(\epsilon_{P}^{3}).
\end{equation}
The background (two unboosted black holes) Christoffel symbol is
\begin{equation}
\label{app_Christoffel_adds}
\Gamma_{jk}^{i(0)} = \Gamma_{jk}^{i(0)C_{1}} + \Gamma_{jk}^{i(0)C_{2}} +{\mathcal O}(\epsilon_{P}^{2}),
\end{equation}
where $\Gamma_{jk}^{i(0)C_{2}}$ is ${\mathcal O}(\epsilon_{P}^{4})$ at a distance $s/2$ or more away from hole $2$, and $\Gamma_{jk}^{i(0)C_{1}}$ has the analogous behavior.  Together, equations (\ref{app_Kij_adds}) and (\ref{app_Christoffel_adds}) imply that the covariant derivative of the extrinsic curvature is
\begin{equation}
D_{k}K_{ij} = D^{C_{1}}_{k}K_{ij}^{C_{1}P_{1}} + D^{C_{2}}_{k}K_{ij}^{C_{2}P_{2}} + {\mathcal O}(\epsilon_{P}^{3}),
\end{equation}
where the covariant derivative in the first term on the right hand side takes into account metric terms from only the first black hole, and the second term is similar.  From equation (\ref{B_ij_Defined}) the magnetic part of the Weyl tensor is therefore equation (\ref{Bijaddsapp}), as desired.

\subsection{The Electric Part of the Weyl Tensor}
\label{APP_CALCS_BINARY_ELECTRIC}

Our calculations for ${\mathcal E}_{ij}$ near hole $1$ are very similar to those for the single boosted black hole in Appendix \ref{APP_CALCS_SINGLEBOOSTED}.  Using equation (\ref{virial}) the conformal factor (\ref{pertbinpsisol}) can be expanded as in equation (\ref{conformal_single_boosted_short}), with the difference being that the perturbation term now has a contribution due to the presence of hole $2$ in addition to the term due to the boost of hole $1$
\begin{eqnarray}
\label{app_binarypsipert}
\epsilon_{P}^{2}u + {\mathcal O}(\epsilon_{P}^{4}) &=& \frac{{\mathcal M}}{2s} + \epsilon_{P}^{2}u_{1}(r) + {\mathcal O}(\epsilon_{P}^{4}) \nonumber\\ &=&\epsilon_{P}^{2}\left(1+u_{1}(r)\right) + {\mathcal O}(\epsilon_{P}^{4}).
\end{eqnarray}
It was shown in \cite{DenBP06} that the effect of the boost of hole $2$ is of higher order.  To lowest order, the extrinsic curvature near hole $1$ is also unaffected by hole $2$, as we discussed above.  We can use the same approach as in Appendix \ref{APP_CALCS_SINGLEBOOSTED} to find the nonzero components of the perturbation of ${\mathcal E}_{ij}$ to second order in $\epsilon_{P}^{2}$, except that we now replace $u$ with $1 + u$.  This calculation yields the nonzero components
\begin{eqnarray}
\label{E_rr_P2_binary}
{\mathcal E}_{rr}^{(2)} &=& \frac{{\mathcal M}}{20 r^4 ({\mathcal M}+2 r)^8} \Bigg(3 {\mathcal M} \bigg(-2 r \Big(\cos 2 \theta \Big) \Big(21 {\mathcal M}^7\nonumber\\&&+\:399 {\mathcal M}^6 r+3052 {\mathcal M}^5 r^2+12194 {\mathcal M}^4 r^3\nonumber\\&&+\:27344 {\mathcal M}^3 r^4+33712 {\mathcal M}^2 r^5+19776 {\mathcal M} r^6\nonumber\\&&+\:3360 r^7\Big)-7
   {\mathcal M} \Big(3 (\cos 2 \theta )+1\Big) \Big({\mathcal M}+4 r\Big)\nonumber\\&&\times\:\Big({\mathcal M}+6 r\Big) \Big({\mathcal M}+2 r\Big)^5 \ln \Big(\frac{{\mathcal M}}{{\mathcal M}+2 r}\Big)\bigg)\nonumber\\&&-\:2 r \bigg(21 {\mathcal M}^8+399 {\mathcal M}^7 r+3142 {\mathcal M}^6 r^2\nonumber\\&&+\:12914 {\mathcal M}^5 r^3+29144 {\mathcal M}^4
   r^4+33712 {\mathcal M}^3 r^5\nonumber\\&&+\:12576 {\mathcal M}^2 r^6-8160 {\mathcal M} r^7-1920 r^8\bigg)\Bigg),
\end{eqnarray}
\begin{eqnarray}
\label{E_rtheta_P2_binary}
{\mathcal E}_{r\theta}^{(2)} &=& \frac{3 {\mathcal M}^2}{20 r^3 ({\mathcal M}+2 r)^8} \Bigg(3840 r^8 \sin \theta  \cos \theta\nonumber\\&&-\:\bigg(\sin 2 \theta \bigg) \bigg({\mathcal M}+2 r\bigg) \bigg(2 r \Big(21 {\mathcal M}^6+357 {\mathcal M}^5 r\nonumber\\&&+\:2170 {\mathcal M}^4 r^2+6342 {\mathcal M}^3 r^3+9388 {\mathcal M}^2
   r^4\nonumber\\&&+\:6216 {\mathcal M} r^5+720 r^6\Big)+21 {\mathcal M} \Big({\mathcal M}+8 r\Big) \nonumber\\&&\times\:\Big({\mathcal M}+2 r\Big)^5 \ln \Big(\frac{{\mathcal M}}{{\mathcal M}+2 r}\Big)\bigg)\Bigg),
\end{eqnarray}
\begin{eqnarray}
\label{E_thetatheta_P2_binary}
{\mathcal E}_{\theta\theta}^{(2)} &=& \frac{{\mathcal M}}{20 r^2 ({\mathcal M}+2 r)^8}\Bigg(3 {\mathcal M} \bigg(7 {\mathcal M} \Big({\mathcal M}+2 r\Big)^5 \nonumber\\&&\times\:\ln \Big(\frac{{\mathcal M}}{{\mathcal M}+2 r}\Big) \Big(3 \big(\cos 2 \theta \big) \big({\mathcal M}^2+7 {\mathcal M} r\nonumber\\&&+\:14 r^2\big)+\big(2 r-{\mathcal M}\big) \big({\mathcal M}+3 r\big)\Big)\nonumber\\&&+\:2 r
   \Big(\cos 2 \theta \Big) \Big( 21 {\mathcal M}^7+336 {\mathcal M}^6 r+2275 {\mathcal M}^5 r^2\nonumber\\&&+\:8330 {\mathcal M}^4 r^3+17528 {\mathcal M}^3 r^4+20608 {\mathcal M}^2 r^5\nonumber\\&&+\:11664 {\mathcal M} r^6+2400 r^7\Big)\bigg)\nonumber\\&&-\:2 r \bigg({\mathcal M}-2 r\bigg) \bigg(21 {\mathcal M}^7+252 {\mathcal M}^6
   r\nonumber\\&&+\:1180 {\mathcal M}^5 r^2+2602 {\mathcal M}^4 r^3+2200 {\mathcal M}^3 r^4\nonumber\\&&-\:1200 {\mathcal M}^2 r^5-3360 {\mathcal M} r^6-480 r^7\bigg)\Bigg),
\end{eqnarray}
and
\begin{eqnarray}
\label{E_phiphi_P2_binary}
{\mathcal E}_{\phi\phi}^{(2)} &=& \frac{{\mathcal M} (\sin ^2\theta ) }{20 r^2 ({\mathcal M}+2 r)^8}\Bigg(3 {\mathcal M} \bigg(7 {\mathcal M} \Big({\mathcal M}+2 r\Big)^5 \times\nonumber\\&&\ln \Big(\frac{{\mathcal M}}{{\mathcal M}+2 r}\Big) \Big(2 {\mathcal M}^2+3 r \big(\cos 2 \theta \big) \big(3 {\mathcal M}+10 r\big)\nonumber\\&&+\:11 {\mathcal M} r+18
   r^2\Big)+6 r^2 \Big(\cos 2 \theta \Big) \Big(21 {\mathcal M}^6\nonumber\\&&+\:259 {\mathcal M}^5 r+1288 {\mathcal M}^4 r^2+3272 {\mathcal M}^3 r^3\nonumber\\&&+\:4368 {\mathcal M}^2 r^4+2704 {\mathcal M} r^5+320 r^6\Big)\bigg)\nonumber\\&&+\:2 r \bigg(42 {\mathcal M}^8+609 {\mathcal M}^7 r+3818
   {\mathcal M}^6 r^2\nonumber\\&&+\:13156 {\mathcal M}^5 r^3+26140 {\mathcal M}^4 r^4+28112 {\mathcal M}^3 r^5\nonumber\\&&+\:11616 {\mathcal M}^2 r^6-1920 {\mathcal M} r^7-960 r^8\bigg)\Bigg).
\end{eqnarray}
\end{appendix}



\end{document}